\UseRawInputEncoding

\documentclass[galaxies,article,accept,pdftex,moreauthors]{mdpi} 
\usepackage{amsfonts}
\usepackage{amssymb}
\usepackage{amsmath}
\usepackage{wasysym}
\usepackage{natbib}
\usepackage{stmaryrd}

\usepackage{graphicx}
\usepackage{moreenum}
\usepackage{rotating}
\usepackage{xcolor}
\usepackage{setspace}
\usepackage[most]{tcolorbox}
\usetikzlibrary{patterns}
\pgfdeclarepatternformonly{mystrikeout}{\pgfqpoint{-1pt}{-1pt}}{\pgfqpoint{11pt}{11pt}}{\pgfqpoint{10pt}{10pt}}%
{
  \pgfsetlinewidth{0.4pt}
  \pgfpathmoveto{\pgfqpoint{0pt}{0pt}}
  \pgfpathlineto{\pgfqpoint{10.1pt}{10.1pt}}
  \pgfusepath{stroke}
}
\newtcolorbox{tcbstrikeout}{breakable,
 enhanced jigsaw,
 opacityback=0,
 parbox=false,
 boxrule=0mm,
 top=0mm,bottom=0pt,left=0pt,right=0pt,
 boxsep=0pt,
 frame hidden,
 finish={\fill[pattern=mystrikeout] (frame.north west) rectangle (frame.south east);}
}
\newcommand{\be}{\begin{equation}}
\newcommand{\ee}{\end{equation}}

\def\Mvir{M_{\rm vir}}
\def\Rvir{R_{\rm vir}}

\def\vir{{\rm vir}}

\def\aap{Astron. Astrophys.}
\def\aj{Astron. J.}
\def\apj{Astrophys. J.}
\def\apjl{Astrophys. J. Lett.}
\def\apjs{Astrophys. J. Suppl.}
\def\apss{Astrophys. Space Sci.}
\def\jcap{J. Cosmol. Astropart. Phys.}
\def\mnras{Mon. Not. R. Astron. Soc.}
\def\nat{Nature}
\def\na{New Astron.}

\def\pasp{Publ. Astron. Soc. Pac.}
\def\physrep{Phys. Rep.}
\def\prd{Phys. Rev. D}

\firstpage{1} 
\pubvolume{1}
\issuenum{1}
\articlenumber{0}
\pubyear{2022}
\copyrightyear{2022}
\externaleditor{Academic Editor: Orlando Luongo and Hernando Quevedo 
}
\datereceived{31 March 2022} 
\dateaccepted{11 May 2022} 
\datepublished{} 
\hreflink{https://doi.org/} 

\Title{A Mass Dependent Density Profile from Dwarfs to Clusters}

\TitleCitation{A Mass Dependent Density Profile from Dwarfs to Clusters}

\Author{Antonino Del~Popolo
 $^{1,2,\dagger}$ 
and Morgan~Le~Delliou $^{3,4,5,}$*$^{,\dagger}$ 
}

\AuthorNames{Antonino Del Popolo, Firstname Lastname and Morgan~Le~Delliou}

\AuthorCitation{Del Popolo, A.; Le~Delliou, M.}

\address{%
$^{1}$ \quad Dipartimento di Fisica e Astronomia, University Of Catania, Viale Andrea Doria 6, 95125 Catania, Italy; adelpopolo@oact.inaf.it\\
$^{2}$ \quad Institute of Astronomy, Russian Academy of Sciences, 119017, Pyatnitskaya str., 48 , Moscow, Russia\\
$^{3}$ \quad 
Institute 
 of Theoretical Physics, School of Physical Science and Technology, Lanzhou University, No.222, South Tianshui Road, Lanzhou 730000 
, China\\
$^{4}$ \quad Instituto de Astrof\'isica e Ci\^encias do Espa\c co, Universidade de Lisboa, Faculdade de Ci\^encias, Ed. C8, Campo Grande, 1769-016 Lisboa, Portugal\\
$^{5}$ \quad Lanzhou Center for Theoretical Physics, Key Laboratory of Theoretical Physics of Gansu Province, Lanzhou University, Lanzhou 730000, China
}

\corres{Correspondence: delliou@lzu.edu.cn or Morgan.LeDelliou.ift@gmail.com 
}

\firstnote{These authors contributed equally to this work.}

\abstract{In this paper, we extend the work of Freundlich {\it et al. }2020 
 who 
{showed how to obtain a }Dekel--Zhao density profile 
{with mass dependent shape parameters in the case of galaxies. }In the case of Freundlich {\it et al. }2020
, the baryonic dependence was obtained using the NIHAO set of simulations. In our case, we used simulations based on a model of ours. Following Freundlich {\it et al. }2020
, we obtained the dependence from baryon physics of the two shape parameters, obtaining in this way a mass dependent Dekel--Zhao profile describing the dark matter profiles from galaxies to clusters of galaxies. The extension to the Dekel--Zhao mass dependent profile to clusters of galaxies is the main result of the paper. In the paper, we show how the Dekel--Zhao mass dependent profile gives a good description of the density profiles of galaxies, already shown by Freundlich {\it et al. }2020
, but also to a set of clusters of galaxies. 
}


\begin{document}

\section{Introduction}

As shown by a series of observations and gravitational effects at the cosmological level~\citep{Planck2016}, together with effects at astrophysical scales 
~\citep{Bertone2005,DelPopolo2014}, we 
know that the Universe cannot only be constituted 
{of }
baryonic matter but should 
contain a non-baryonic mass/energy component with the property of clustering {and which is }dubbed dark matter (DM). Similarly, the galaxies appear 
constituted by an amount of baryons that make them visible, and a large halo of DM in which the baryonic component is embedded. {Many studies }
have been performed to describe the DM halo density profiles. Some {have }
showed that, from dwarf {galaxy }haloes to {galaxy }clusters, the density profile {can }
be described by the 
Navarro--Frenk--White (NFW) profile
\citep{NFW1996,NFW1997}, characterized by a double power law at small masses ($\rho \propto r^{-1}$), and large masses ($\rho \propto r^{-3}$). Further studies found a different kind of profile whose 
slope decreases going toward the center of the halo, e.g.,~\citep{Navarro2004,Navarro2010,Gao2008,Springel2008}. These kinds of profiles are well described by the so-called Einasto profile of which we will speak later. 
The cuspy behavior of the NFW in the central region of the halo disagrees with 
observations {of}
low-surface-brightness{ (LSB)}, dwarf satellite galaxies, and even {to a small extent with galaxy }clusters, which show flatter cores, e.g.,~\citep{Flores1994,Moore1994,deBlok2008,deBlok2010,Kuzio2011,Oh2015,Newman2013a,Newman2013b,Adams2014}. 
In other words, the smallest of the inner slope of the density profile
value predicted by dissipationless N-body simulations is larger than the values 
obtained by observations 
\citep{Burkert1995,deBlok2003,Swaters2003,KuziodeNaray2011,Oh2011a,Oh2011b}, in SPH {(Smooth Particle Hydrodynamics) }simulations 
\citep{Governato2010,Governato2012}, or in semi-analytical models 
\citep{DelPopolo2009,Cardone2012,DelPopolo2012a,DelPopolo2012b,DelPopolo2014a}. 
 
This issue {presents }
problems {for }
the $\Lambda$CDM model at small scales and is called the Cusp/Core problem~\citep{Moore1994,Flores1994} (other problems often mentioned are 
 (a) the missing satellite problem, namely the discrepancy between the number of subhaloes that N-body simulations predict, e.g.,~\citep{Moore1999} 
and observations; (b) and the Too-Big-To-Fail (TBTF) problem, characterized by too many and too dense simulated sub-haloes with respect to observations \citep*{BoylanKolchin2011,BoylanKolchin2012}). 
 
As mentioned, apart from the case of dwarf galaxies and LSBs, the cusp/core problem is also present at the scales of clusters of galaxies, mainly in their central part 
(10 kpc \cite{Newman2013a,Newman2013b}). 

As found by kinematics and lensing constraints in galaxies located in the center of relaxed clusters (cD galaxies, Brightest Central Galaxy, hereafter BCG), while the total mass profile, namely the sum of DM 
and baryons, is in agreement with the NFW predictions, the cluster{'}s DM profile 
is flatter than a{n} NFW profile~\citep{Sand2002,Sand2004,Newman2013a,Newman2013b}.
 
Several ideas have been proposed to solve that 
discrepancy: {broadly, they are }cosmological {or }
astrophysical. 

The first approach is based on cosmological solutions{ to the Cusp/Core problem}, namely different DM types of 
particles are considered rather than, for example, the WIMPS (Weakly Interacting Massive Particles)
\citep{Colin2000,Goodman2000,Hu2000,Kaplinghat2000,Peebles2000,SommerLarsen2001}. Some solutions 
introduced modification in the spectrum at small scales e.g.,~\citep{Zentner2003}. Other solutions 
modified theories of gravity 
used, such as  
$f(R)$~\citep{Buchdahl1970,Starobinsky1980}, $f(T)$ 
see~\citep{Bengochea2009,Linder2010,Dent2011,Zheng2011} and MOND {(MOdified Newtonian Dynamics)}~\citep{Milgrom1983b,Milgrom1983a}. 

The main idea at the root of 
astrophysics solutions resides in 
the existence of a ``heating mechanism'' that produces an expansion of the DM component of a galaxy. This, in turn, produces a reduction of the inner density. In the ``supernovae feedback flattening'' mechanism, supernovae explosions and/or 
AGN {(Active Galactic Nucleus) }outflows 
produce potential fluctuations that can heat the DM, producing a flattening of the cusp 
\citep{Navarro1996a,Gelato1999,Read2005,Mashchenko2006,Mashchenko2008,Governato2010,Governato2012}.

A different mechanism is based on the exchange of angular momentum and energy between infalling gas clumps, and DM, with the result of ``heating'' {of the }DM~\citep{ElZant2001,ElZant2004,Ma2004,Nipoti2004,RomanoDiaz2008,RomanoDiaz2009,DelPopolo2009,Cole2011,Inoue2011,
Nipoti2015}.
 
The NFW
, 
Einasto, 
{and related }
density profiles are 
independent{ of the mass scale}. In~\citep{DelPopolo2010}{, it}  
was shown that the inner slope {of the density profile }depends {on }
the halo mass. This result was later confirmed by several simulations both in the case of galaxies and clusters
e.g.,~\citep{DiCintio2014,Zolotov2012,Martizzi2013,Teyssier2013,Chan2015, Tollet2016, Peirani2017,Maccio2020}. 

Some papers~\citep{DiCintio2014,Chan2015,Tollet2016,Dutton2016} showed that 
the inner slope in SPH simulations displays 
regions with values similar to or larger than the NFW slope, and others that are much flatter, reaching 
a minimum for masses down to 
$10^{8.5}\rm M_\odot$, in the case of~\citep{DiCintio2014}.
This minimum is due to the fact that {mass }outflows {of explosive events }overcome halo gravity, giving rise to an expansion of the halo. For larger masses, the slope steepens and reaches values larger than 
the NFW slope when the stellar mass exceeds $10^{10}~\rm M_\odot$. At those masses, baryon 
accumulation gives rise to adiabatic contraction. The {latter effect }
can be counteracted in {the }presence of AGN feedback~\citep{Maccio2020,DelPopolo2021}.

In order to fit the density profiles, several parameterizations have been used. The 
Einasto profile~\citep{Einasto1965, Navarro2004,Mamon2010, Retana-Montenegro2012, An2013} presents 
two free shape parameters. Leaving free in the fit these two parameters, one can get good fits to DM cusps. Moreover, analytic expressions for the mass, the surface density, the gravitational potential, and quantities important for lensing can be 
provided. Unfortunately, the Einasto profile does not give a good fit to the innermost part of the density profile {for galaxies }\citep{Dekel2017}. Another profile often used is the generalized 
NFW (gNFW) model that performs in a similar way to the Einasto profile. Other density profiles used have adopted the averaged 
form.
\begin{equation}
\label{eq:rho_abc}
\rho(r) = \frac{\rho_c}{x^a (1+x^{1/b})^{b(g-a)}}
\end{equation}
as in \citet{Zhao1996},
{where }
$x=r/r_c$, with $r_c$, a characteristic radius and $\rho_c$, a typical density Note that the non-averaged DZ density profile coincides with gNFW profiles for the choice $g=3,b=2$).
This is also the functional form of the gNFW density profile. Considering this form for the actual density profile, if 
$b=n$ and $g=3+k/n$, where $n$ and $k$ can be any natural numbers, and one can obtain analytically the gravitational potential, the mass, and the velocity dispersion.
This 
family of density profiles is able to provide one which yields 
very good fits to DM simulations even when baryons are present and is able to follow the profile even when cusps or cores are present. This profile, as shown by \citet{Dekel2017}, corresponds to 
$n=2$ and $k=1$, i.e., $b=2$ and $g=3.5$ in Equation~(\ref{eq:rho_abc}). It remains with two free 
parameters, namely, $a$ and $c$, is usually referred to as the 
Dekel--Zhao (DZ) profile, and captures cores better than other profiles~\citep{Freundlich2020}. In order to improve the fit to the density profiles of galaxies and clusters, a new approach has been introduced. The parameters of each fitting profile 
are expressed in terms of stellar mass, $M_{*}$, and virial mass, $M_{\rm vir}$, or their ratio. For example, Ref.~\cite{Lazar2020} proposed a modified Einasto profile, and Ref.~\cite{Read2016} a modified NFW. Unfortunately, their results are obtained at the expense of analyticity. Similarly, Ref.~\cite{DiCintio2014b} obtained a density profile giving the functional form of the parameters $a, b, g$ and the concentration parameter, $c$, in terms of 
$M_{\star}/M_{\rm vir}$. Their result is a mass-dependent density profile
(\citep{DiCintio2014b}~hereafter Di Cintio {et~al.}) for the haloes. Their {corresponding }gravitational potential, 
velocity dispersion, and lensing properties do not {possess }
any analytic form{s}.

\textls[-5]{Recently, a mass dependent profile based on the DZ profile was obtained by~\citet{Freundlich2020}. Similarly to \citet{DiCintio2014b}, they used the NIHAO project to obtain the functional form of the two parameters of the DZ density profile. The validity of~\citet{Freundlich2020},~\citet{DiCintio2014b}}, and all the mass dependent profiles obtained in the literature is limited to 
galaxies because the simulations at their base do not take account of AGN feedback. Therefore, their 
mass dependent profile cannot be built up to 
the clusters of galaxy mass range. In this paper, our goal is to extend the DZ profile to clusters of galaxies.

The NIHAO project is based on SPH simulations, whose 
core formation base mechanism focuses on 
``supernovae feedback''. We previously mentioned the alternat{iv}e 
core formation mechanism based on the exchange of energy and angular momentum between infalling gas clumps and DM. We used it 
in several papers, and describe the model 
in the next section. Generating 
structures of different size with it, we {seek }
in this paper to 
find a relation between the parameters of the DZ density profile, from galaxies to clusters. 
The paper is organized as follows:
{In Section~\ref{sec:Implementation}. we describe the model used. In Section~\ref{sec:outlineModel}, we summarize the steps in which the model works. In Section~\ref{sec:DZ}, we describe the DZ profile in general and in Section~\ref{sec:massDepDZ} the DZ mass dependent profile. In Section~\ref{sec:clustersDen}, we apply the previous profile to clusters of galaxies. Section~\ref{sec:conclusions} is devoted to conclusions.

}

\section{Description of the Semi-Analytic Model}
\label{sec:Implementation}
{We }
employ 
the semi-analytic model of \cite[][]{DelPopolo2009,DelPopolo2009a}. This scheme significantly improves on the classical spherical collapse models~\citep{Gunn1972,Bertschinger1985,Hoffman1985,Ryden1987,Ascasibar2004,Williams2004} by taking into account the impact of
\begin{enumerate}[leftmargin=*,label={\tiny$\diamond
$},noitemsep,partopsep=0pt,topsep=0pt,parsep=0pt]
\item tidal torque 
  induced ordered angular momentum  e.g.,~\citep{Ryden1988,DelPopolo1997,DelPopolo2000},
\item random angular momentum caused by {random motion in the halo collapse phase }
  e.g.,~\citep{Ryden1987,Williams2004}, 
 \item adiabatic contraction AC, e.g.,~\citep{Blumenthal1986,Gnedin2004, Klypin2002,Gustafsson2006}, 
 \item DM's dynamical friction with baryonic stellar clumps and gas~\citep{ElZant2001,ElZant2004,Ma2004,RomanoDiaz2008,RomanoDiaz2009,DelPopolo2009,Cole2011,Inoue2011, Nipoti2015}, 
 \item feedback from baryonic-radiation interactions, such as gas cooling, star formation, photoionization, supernova, and AGN feedback~\citep{DeLucia2008,Li2010,Martizzi2012} and  
 \item {dark energy (}DE{), described by the cosmological constant, as can be seen in \mbox{Equation A14} in \cite{DelPopolo2009}, giving the equation of motion of a generic shell}~\citep{DelPopolo2013a,DelPopolo2013b,DelPopolo2013c}, 
\end{enumerate}
and by further refinements \cite{DelPopolo2014a,DelPopolo2016a,DelPopolo2016b,DelPopolo2016b}. These enhancements allowed for accurate and improved results on
\begin{enumerate}[leftmargin=*,label={\tiny$\Yleft
$},noitemsep,partopsep=0pt,topsep=0pt,parsep=0pt] 
 \item density profiles universality~\citep{DelPopolo2010,DelPopolo2011},
 \item distinct density profiles details in
 \begin{enumerate}[leftmargin=*,label={\tiny$\rightslice$},noitemsep,partopsep=0pt,topsep=0pt,parsep=0pt]
  \item galaxies~\citep{DelPopolo2012a,DelPopolo2014} and
  \item clusters~\citep{DelPopolo2012b,DelPopolo2014},
 \end{enumerate}
\item inner {galactic }
  surface-density \citep*{DelPopolo2013d}.
\end{enumerate}

While focused on the dynamical friction mechanism (DFBC), the semi-analytic scheme includes contributions from each of the effects listed above, including supernova explosions feedback (SNF) that each impact results to the order of a few {percent}
.

The scheme is implemented according to the following stages:
\begin{enumerate}[leftmargin=*,label=(\roman*),noitemsep,partopsep=0pt,topsep=0pt,parsep=0pt]
 \item Expansion of the diffuse gas and DM proto-structure in the linear phase, reaching a maximum radius before re-collapse of DM, forming a potential well for baryons to fall;
\item Formation of stars from baryons radiative clumping in the halo center;
\item Four {parallel }
  processes then follow;
\begin{enumerate}[leftmargin=*,noitemsep,partopsep=0pt,topsep=0pt,parsep=0pt]
\item baryon 
  AC increases the DM central cusp, e.g., for {$10^9 M_{\odot}$} galaxies, at {$z \simeq 5$}, see~\citep{DelPopolo2009},

\item baryon
  -DM dynamical friction (DF) collapses clumps to the galactic center,
 \item the halo central density reduces~\citep{ElZant2001,ElZant2004} from the transfer to DM, and stars~\citep{Read2005,Pontzen2012,Teyssier2013} of DF energy and angular momentum (AM), contrary to the effect of AC;
 \item DF and AC balance, opening the possibility for cusps to heat up, some up to core formation, as in spirals and dwarf spheroidals, while others, like giant galaxies, retains their steeper profile and cusp because of their deeper potential wells;
\end{enumerate}
\item Tidal torques (ordered AM), and random AM join their similar effects to DF. 
  \item \label{SNmech} Finally, the disruption of the smallest gas clumps, due to their partial conversion to stars, and the supernovae explosions repeated gas expulsion decrease stellar density, resulting in an additional slight core enlargement; see~\citep{Nipoti2015}.
\end{enumerate}

\subsection{Density Profile Generation}

We model the emergence of the density profile in the framework of the spherical model. From an initial, linear Hubble expansion, the expanding density perturbations eventually reaches a maximum turn-around, marking the onset of its recollapse~\citep{Gunn1977,Fillmore1984}. A particle-based Lagrangian approach is used to compute the final density profile, noting the particle's initial and turn-around radii $x_{\rm i}$ and $x_{\rm m}(x_{\rm i})$, its turnaround density $\rho_{\rm ta}(x_{\rm m})$ and its collapse factor $f(x_{\rm i})=x/x_{\rm m}(x_{\rm i})$, in order to obtain
\begin{equation}\label{eq:dturnnn}
 \rho(x)=\frac{\rho_{\rm ta}(x_{\rm m})}{f(x_{\rm i})^3}
 \left[1+\frac{d\ln{f(x_{\rm i})}}{d\ln{g(x_{\rm i})}}\right]^{-1}\;.
\end{equation}
In this approach, one computes the turn-around radius from the density parameter $\Omega_{\rm i}$ and the
DM and baryon 
shell's average overdensity $\overline{\delta}_{\rm i}$
\begin{equation}
 x_{\rm m}=g(x_{\rm i})=x_{\rm i}\frac{1+\overline{\delta}_{\rm i}}
 {\overline{\delta}_{\rm i}-(\Omega_{\rm i}^{-1}-1)}\;. 
\end{equation}

Baryons start entirely in gas form, denoted by the ``universal baryon fraction''~\citep{Komatsu2009}, set to 0.167 in \citep {Komatsu2011} over the total mass and set to $f_{\rm b}=0.17\pm 0.01$, before stars form as discussed below (Sections~\ref{sec:clumpLifeTime} and \ref{sec:FeedbackSF}). 

The profile is then modified by the effects of ``specific ordered angular momentum'', $h$, computed from Tidal torque theory (TTT) that describes how larger scales' tidal torques induce smaller scales' angular momentum~\citep{Hoyle1953,Peebles1969,White1984,Ryden1988,Eisenstein1995}, as well as from ``random angular momentum'', $j$, which can be computed from particle orbits, specified by their orbit{al }
eccentricity{, encoded in the ratio} $e=\left(\frac{r_{\rm min}}{r_{\rm max}}\right)$~\citep{AvilaReese1998}, using {their }
pericentric and apocentric radii, resp. $r_{\rm min}$ and $r_{\rm max}$. The dynamical state of the system induces a correction to the eccentricity \cite{Ascasibar2004}, computed from the halo maximum and spherically averaged turnaround radii, $r_{\rm ta}=x_{\rm m}(x_{\rm i})$ and $r_{\rm max}<0.1 r_{\rm ta}$, as
\begin{equation}
e(r_{\rm max})\simeq 0.8\left(\frac{r_{\rm max}}{r_{\rm ta}}\right)^{0.1}\;.
\end{equation}

These effects, as well as DFs, as described by the DF force equation of motion, see Equation A14 in~\citep{DelPopolo2009}, and AC's steepening, following \cite{Gnedin2004}, result in the final density profile.

\subsection{Inclusion of the Baryonic Discs and Clumps Effects}

In spiral galaxies, the baryon gas halo evolves into a rotationally supported, stable disk, following the equation of motion to {obtain a }
realistic disc size and mass that also solves the angular momentum catastrophe (AMC){ problem}, Section 3.2, Figures 3 and 4 of~\citep{DelPopolo2014}.

\subsubsection{Clump 
  Size Calculation}

When disks grow denser, instability appears {due to }
Jean's criterion in spite of shear effect 
stabilization. This instability leads to clump formation, whose condition was described by \cite{Toomre1964}, defining a limit criterion composed with $\sigma$ ($\simeq$20--80 km/s, for most galaxies hosting a clump),
 the 1D disk's velocity dispersion, $\Sigma$, its surface density, connected to its adiabatic sound speed $c_s$, $\Omega$, its angular velocity and $\kappa$, its epicyclic frequency 
\begin{equation}
Q \simeq \sigma \Omega/(\pi G \Sigma)=\frac{c_s \kappa}{\pi G \Sigma}<1\;.
\end{equation}
The dispersion relation for perturbations, $d \omega^2/d k=0$, yields solutions with the mode that grows fastest \cite{BinneyTremaine1987}
\[k_{\rm inst}=\frac{\pi G \Sigma}{c_s^2},\] for $Q<1$ (Equation 6 in \cite{Nipoti2015}) that leads to obtain the galaxies clump radii~\citep{Krumholz2010}
\begin{equation}
 R \simeq 7 G \Sigma/\Omega^2 \simeq 1 {\rm kpc}\;.
\end{equation}
The total mass of maximal{--}%
velocity{--}%
dispersion, marginally unstable discs ($Q \simeq 1$) reaches 3 
times that of the cold disc and can convert $\simeq 10$ \% of their disk mass $M_d$ into clumps~\citep{Dekel2009}.

{The }
main properties {of clumps }found by \cite{Ceverino2012} reflect the general case. For instance, {at }$z \simeq 2$, $5 \times 10^{11} M_{\odot}$ haloes harbor objects of order $10^{10}~M_{\odot}$ that remain marginally unstable for $\simeq 1$~Gyr. 

Smaller haloes have their profiles more efficiently flattened by clumps to DM transfer of AM and energy, as found by \cite{Ma2004,Nipoti2004,RomanoDiaz2008,RomanoDiaz2009,DelPopolo2009,Cole2011,Inoue2011,DelPopolo2014d,Nipoti2015}. 

\subsubsection{Clump 
  Life-Time Calculation\label{sec:clumpLifeTime}}

Clumps are proven to exist in both simulations, e.g., \citep
{Ceverino2010,Perez2013,Perret2013,Ceverino2013,Ceverino2014,Bournaud2014,Behrendt2016} and observations, as, for instance, the {clusters of clumps }
or clumpy structures detected in high redshift galaxies, thus dubbed chain galaxies e.g.,~\citep{Elmegreen2004,Elmegreen2009,Genzel2011}, or the massive star-forming clumps found in HST {(Hubble Space Telescope) }Ultra Deep Field galaxies~\citep{Guo2012,Wuyts2013}, {of which several have been }
observed at lower redshift $z=1-3$~\citep{Guo2015}, {and }a few at deeper redshift $z \apprle 6$ \cite{Elmegreen2007}.

Very gas-rich discs are expected to give birth, through self-gravitation instability from the accreting dense gas radiative cooling, to those clumpy structures e.g.,~\citep{Noguchi1998,Noguchi1999,Aumer2010,Ceverino2010,Ceverino2012}. Clump lifetime is central to their impact on halo central density: they can flatten a cusp into a core if they survive stellar feedback disruption long enough to sink to the galactic center. The assessment of a stellar clump's bound state can be obtained through $e_f$, its mass fraction loss by stellar feedback, and $\varepsilon=1-e_f$, its stellar mass fraction. Then, a threshold at $\varepsilon \geq 0.5$ marks the limit for most of the stellar clump's mass to remain bound, as confirmed by simulations and analytical models \cite{Baumgardt2007}. The efficiency of stellar radiation feedback is evaluated from the expulsion fraction $e_f=1-\varepsilon=0.086 (\Sigma_1 M_9)^{-1/4} \epsilon_{eff},_{-2}$~\citep{Krumholz2010}, with
\begin{enumerate}[labelsep=1pt,
leftmargin=21pt,
label=(\itshape\roman*\upshape),noitemsep,partopsep=0pt,topsep=0pt,parsep=0pt]
\item 
$\Sigma_1=\frac{\Sigma}{0.1 g/cm^2}$, the dimensionless reduced surface density 
\item 
$M_9=M/10^9 M_{\odot}$, the dimensionless reduced mass and
\item  
$\epsilon_{eff},_{-2}=\epsilon_{eff}/0.01$, the dimensionless reduced efficiency rate of the star-formation, with $\epsilon_{eff}=\frac{\dot{M_*}}{M/t_{ff}}$ simply obtained from the ratio between free-fall time, $t_{ff}$ and the stellar mass $M_\star$ depletion time.
\end{enumerate}

A large sample of densities, sizes, environments, and scales were used in Ref.~\citep{Krumholz2007} to obtain $\epsilon_{eff}\simeq 0.01$.
As typical {clumps, having masses }$M\simeq10^9 M_{\odot}${, exhibit }
$e_f=0.15$ and $\varepsilon=0.85$, their mass loss after reaching the galactic halo center should remain small. The expulsion fraction method and this conclusion are unfortunately valid in smaller galaxies, for smaller, more compact clumps, for which no clump can be disrupted before reaching the center.

An alternat{iv}e method produces clump disruption as a result of a comparison between clump 
migration time to the {structure }center and their lifetime. The latter has been studied extensively. {In their }
hydrodynamical simulations exhibiting rotational supported clumps in Jean's equilibrium, Ceverino et~al. deduced long lifetime{s} ($\simeq$$2 \times 10^8$ Myr) for them, in agreement with several studies: Ref.~\citep{Krumholz2010} found such lifetimes in local systems with Kennicutt--Schmidt law star formation. {Such long lifetimes can be explained by a long enough }
clump 
galactic center migration time {to allow }
them to retain gas, and form bound star groups, as confirmed by simulations from \cite{Elmegreen2008}. Galactic center reaching, long-lived clumps were also produced in other simulations, properly accounting {for }stellar feedback, e.g., radiative and non-thermal feedback, SNF, radiation pressure, etc., in~\citep{Perret2013,Bournaud2014,Ceverino2013}, also presented by \cite{Perez2013}, for any reasonable feedback efficiency. Clump ages estimated through metal enrichment, expansion, and gas expulsion time scales, $\simeq$200 Myr, >100 Myr and 170--1600 Myr, respectively, in \cite{Genzel2011} also favors strongly long-lived clumps. Finally, clump stability also emerges from similar clump 
observations, in radius, mass, and in~\citep{Elmegreen2013,Garland2015,Mandelker2015} between low and high redshifts.

DF and TTT balance off to produce migration time; see Equations 1 and 18 of~\citep{Genzel2011,Nipoti2015}, obtaining $\simeq$200 Myr 
for a $10^9 M_{\odot}$ clump. The Sedov--Taylor solution, Equations 8 and 9 in~\citep{Genzel2011}, produced similar migration and expansion timescales.

\subsection{Feedback and Star Formation Procedure\label{sec:FeedbackSF}}

The model's stellar feedback proceeds from the prescriptions of Sections~2.2.2 and~2.2.3 in \citep{DeLucia2008,Li2010} for gas cooling, reionisation, star formation, SNF and AGN feedback.

\begin{description}
\item[Gas cooling] 
 is modeled with a cooling flow, e.g., see Section 2.2.2 in \citep{White1991,Li2010}.
 \item[Reionisation] decreases the baryon fraction, during the epoch $z=11.5-15$, as
\begin{equation}
f_{\rm b, halo}(z,M_{\rm vir})=\frac{f_{\rm b}}{[1+0.26 M_{\rm F}(z)/M_{\rm vir}]^3}\;,
\end{equation}
Ref. \citep{Li2010}, where $M_{\rm vir}$ is the virial mass and $M_{\rm F}$, the ``filtering mass''; see~\citep{Kravtsov2004}. 
\item[Star formation] occurs when gas converts into stars, after settling in a disk. Over a given time interval $\Delta t$ that can be set to $t_{\rm dyn}$, the disc dynamical time, the amount of gas mass converted into stars can be computed as
\begin{equation}
 \Delta M_{\ast}=\psi\Delta t\;,
\end{equation}
with $\psi$, the star formation rate, obtained from the mass of gas measured at a density above the threshold $n>9.3/{\rm cm^3}$, fixed as in \citep{DiCintio2014} as follows, see \citep{DeLucia2008}, for more details:
\begin{equation}
\psi=0.03 M_{\rm sf}/t_{\rm dyn}\;.
\end{equation}
\item[SNF] explosions inject energy into the halo hot gas, following \cite{Croton2006}. The computation of this injected energy is prescribed from a Chabrier IMF \cite{Chabrier2003}, {consisting of}
  \begin{itemize}[leftmargin=*,label={\tiny$\therefore$},noitemsep,partopsep=0pt,topsep=0pt,parsep=0pt]
                                                           \item $\epsilon_{\rm halo}$, the energy efficiency of disc gas reheating; 
                                                           \item $\Delta M_{\ast}$, the available mass within stars;
                                                           \item $\eta_{\rm SN}=8\times 10^{-3}/M_{\odot}$, the number of SN, created from conversion of $\Delta M_{\ast}$ into SN, per solar mass, and
                                                           \item $E_{\rm SN}=10^{51}$ erg, the typical  energy released per SN explosion, 
                                                          \end{itemize}
 into the total SN injected energy 
\begin{equation}
 \Delta E_{\rm SN}=0.5\epsilon_{\rm halo}\Delta M_{\ast} \eta_{\rm SN}E_{\rm SN}\;.
\end{equation}
This SN released energy into a reheated disk gas and then 
{compared itself} with the reheating energy $\Delta E_{\rm hot}$ which that same amount of gas should acquire if its injection in the halo should keep its specific energy constant, that is, if the new gas would remain at equilibrium with the halo hot gas. The 
amount of disk gas the SN and {stellar }
radiation have reheated, $\Delta M_{\rm reheat}$, 
since it is {all }produced from 
radiation
{ of stellar origin}, is 
proportional to the{ stellar }
mass
\begin{equation}
 \Delta M_{\rm reheat} = 3.5 \Delta M_{\ast}\;.
\end{equation}
Since
the halo hot gas specific energy 
corresponds to the Virial equilibrium specific kinetic energy $\frac{V^2_{\rm vir}}{2}$, keeping this energy constant under the addition of that 
reheated gas leads to defining the equilibrium reheating energy as
\begin{equation}
\Delta E_{\rm hot}= 0.5\Delta M_{\rm reheat} 
V^2_{\rm vir}\;.
\end{equation}
The comparison with the actual energy of the gas injected from the disk into the halo by SNs 
gives the threshold ($\Delta E_{\rm SN}>\Delta E_{\rm hot}$), beyond which gas is expelled, 
the available energy to expel the reheated gas, and thus the amount of gas ejected from that extra energy
\begin{equation}
 \Delta M_{\rm eject}=\frac{\Delta E_{\rm SN}-\Delta E_{\rm hot}}{0.5 V^2_{\rm vir}}\;.
\end{equation}\\
Contrary to SNF based models such as \cite{DiCintio2014}, our mechanism for cusp flattening initiates before the star formation epoch. Since it uses 
a gravitational energy source, it 
is thus less limited in available time and energy. Only after DF shapes the core can Stellar and SN feedback occur
, which then disrupts gas clouds in the core, similarly to~\citep{Nipoti2015}.
\item[AGN feedback] points to the effects and the formation of a central Super-Massive-Black-Hole (SMBH). Our approach adopts the SMBH mass accretion, and subsequent AGN feedback models of \citet{Booth2009}, modified by the \citet{Martizzi2012,Martizzi2012a} prescriptions. When the thresholds $2.4 \times 10^6 M_{\odot}/{\rm kpc}^3$, and 100 $\rm km/s$, for stellar density and reduced gas density ($\rho_{gas}/10$), and 3D velocity dispersion, are exceeded, the formation of a seed $10^5~M_{\odot}$ SMBH occurs and it starts accreting. It has been shown~\citep{Cattaneo2006} that, above $M\simeq 6 \times 10^{11} M_{\odot}$, significant AGN quenching occurs. 

\end{description}

\subsection{Confirmations of the Semi-Analytic Model's Robustness}

The semi-analytic model's robustness and accuracy was established in several confrontations with observations and previous converging models:
\begin{enumerate}[label=\greek*.]
\item The{galaxy and cluster }
  cusp flattening predicted by the model from collapsing baryonic clump 
  DM heating agrees with previous studies~\citep{ElZant2001,ElZant2004,RomanoDiaz2008,RomanoDiaz2009,Cole2011,Inoue2011,Nipoti2015}, as shown in its, Figure 4 in \citep{DelPopolo2011} comparison with SPH simulations from \cite{Governato2010};
\item The model predicted{, in advance of competing groups in the field, }
  the halo cusp 
  inner slope mass dependence, Figure 2a solid line in \cite{DelPopolo2010}, in terms of rotation $V_c$, given as $2.8 \times 10^{-2} M_{\rm vir}^{0.316}$~\citep{Klypin2011}, well in advance of the similar conclusion {that one can extrapolate }
  from Figure 6 in \cite{DiCintio2014};
    \item The inner slope dependence on the ratio between baryonic and total mass of the halo was also predicted by the model; see \cite{DelPopolo2012b} well before its more publicized claim \cite{DiCintio2014};
    \item The correct {galaxy }
      density profiles {were }
      also obtained by the model~\citep{DelPopolo2009,DelPopolo2009a} previous to the \cite{Governato2010,Governato2012} SPH simulations, while that for clusters was predicted  in~\citep{DelPopolo2012b}, before the results of \cite{Martizzi2013}. Note that these results from the model were obtained with its different dominant mechanism from those of \cite{Governato2010,Governato2012,Martizzi2013}; 
    \item The Tully--Fisher and Faber--Jackson, $M_{\star}-M_{halo}$, relationships were compared with simulations from the model (Figures 4 and 5 in~\citep{DelPopolo2016a,DelPopolo2016b}).
    \item Finally, the model's inner slope evolution with mass (Figure 1 in~\citep{DelPopolo2016a,DelPopolo2016b}) agrees with~\citep{DiCintio2014}'s simulations.
\end{enumerate}

\section{Outline of the Semi-Analytic Model's Main Steps}\label{sec:outlineModel}

The semi-analytic approach of our model is much less computing intensive than hydrodynamical and/or N-body simulations, such as NIHAO. 
Recall that the NIHAO simulation is based on the GASOLINE2 hydrodynamical simulation that includes effects such as~\citep{Maccio2020} photoionisation, Compton cooling, metal cooling, heating from the ultraviolet background, chemical enrichment, star formation, and feedback from massive stars and from SN.

Our model therefore offers a simpler scheme to construct galaxy samples and a faster parameter space exploration. Semi-analytic and N-body/hydro simulations' comparisons have shown good agreement in the studied cases, see \citep{Benson2012} and references therein. Our cosmological parameters follow, Section 2 in \cite{Maccio2020}, while the system's baryons start in gas form, with ``universal baryon fraction''~\citep{Komatsu2009} set as $f_{\rm b}=0.17\pm 0.01$, set to 0.167 in~\citep{Komatsu2011}. Initial conditions, set from the power spectrum, and evolution follow the description from Appendix B in \citep{DelPopolo2009}. The onset, when the system's nonlinear regime is reached, and subsequent evolution of tidal interaction with neighbours is detailed in Appendix C \citep{DelPopolo2009}. The generation of random angular momentum during the collapse phase is calculated in Appendix D \citep{DelPopolo2009}, while the baryonic dissipative collapse is treated in Appendix E \citep{DelPopolo2009}, when spiral structure or a spheroid shape are generated, as described in Appendix A5 \citep{DelPopolo2016a}. The depiction of the model's clumps characteristics and formation can be found in Section 2.2 in \citep{DelPopolo2016a}, its star formation is specified in Section 2.3 in \citep{DelPopolo2016a}, while its AGN feedback and BH formation are portrayed in Section 2.3, the final part in \citep{DelPopolo2016a}.

\section{The Dekel--Zhao Profile}
\label{sec:DZ}

\textls[-25]{By means of simulated haloes using the NIHAO suite of simulations 
\citep{Wang2015},~\citet{Dekel2017}} showed that the functional form of Equation~(\ref{eq:rho_abc}) with $b=2$ and $g=3.5$ gives very good fits to haloes characterized by cores or 
cusps. As already reported, this parameterization is dubbed DZ. It can be characterized by 
better fits to simulated profiles with respect to the NFW and the Einasto profile, and better captur{e of }
the cores of 
density profiles. Moreover, the DZ parameterization allows for 
analytic expressions of 
the velocity dispersion, the density, the mass, and the gravitational potential.
We now recall several 
important {relations.}

The functional form of the DZ profile is
\begin{equation}
\label{eq:rho32}
\rho(r)  =  \frac{\rho_c}{x^a (1+x^{1/2})^{2(3.5-a)}} 
\end{equation}
with $\rho_c=(1-a/3)\overline{\rho_c}$, $\overline{\rho_c}= c^3 \mu \overline{\rho_\vir}$, $\overline{\rho_{\rm vir}}=3\Mvir/4\pi \Rvir^3$, $\mu = c^{a-3} (1+c^{1/2})^{2(3-a)}$ and $x=r/r_c$, determined by the 
two shape parameters $a$ and $c=\Rvir/r_c$.

The inner logarithmic slope, at 
{scale} 
$r_1$, is given by~\citep{Freundlich2020}
\be
\label{eq:s1_23}
s_1=\frac{a+3.5 c^{1/2}(r_1/\Rvir)^{1/2}}{1+c^{1/2}(r_1/\Rvir)^{1/2}}, 
\ee
and the 
concentration parameter 
smoothed at the same scale, 
by
\be
\label{eq:c2_23}
c_{2}
=c\left(\frac{1.5}{2-a}\right)^2. 
\ee
{The }
condition that the density is positive {yields }
the condition $a\leq 3$, while, from the condition that the inner slope is positive, {we obtain }
the extra 
condition 
$a+3.5c^{1/2}(r_1/\Rvir)^{1/2}\geq 0$. 
%
%
The parameters $(a, c)$ are related to $(s_{\rm 1}, c_{\rm 2})$. Both $a$ and $c$ can be expressed in terms of $s_1$ and $c_{2}$ as 
\be
\label{eq:a(s1,c2)}
a = \frac{1.5 s_1 -2 \left( 3.5-s_1\right)\left(r_1/R_\vir\right)^{1/2}c_{2}^{1/2} }{1.5 - \left(3.5-s_1\right)\left(r_1/R_\vir\right)^{1/2}c_{2}^{1/2} }
\ee
and
\be
\label{eq:c(s1,c2)}
c= \left( \frac{s_1-2}{\left(3.5-s_1\right)\left(r_1/R_\vir \right)^{1/2} -1.5 c_{2}^{-1/2}} \right)^2. 
\ee
As in~\citep{Freundlich2020}, ($a$, $c$) are used to express analytic expressions, and ($s_1$, $c_2$) are used in numerical tests. 

The analytic expression for the gravitational potential, the velocity dispersion, and the lensing properties are given in Equation (18), Equation (21), and Section 2.3 of~\citep{Freundlich2020}, respectively. 

\citet{Freundlich2020} compared the DZ
, the 
gNFW
, and the Einasto profiles using the results of the NIHAO suite of SPH simulations. The authors used some profiles from the NIHAO suite, plotted in their Figure 3, and fitted the NIHAO profiles using the 
gNFW, DZ, and Einasto profiles. The parameters for each profile were considered free. 
\citet{Freundlich2020} showed that the DZ performs better than the other two parameterizations, and in particular, 
significantly better than the Einasto profile, and marginally better than the  
gNFW profile. 

\section{The Dekel--Zhao Mass Dependent Profile}\label{sec:massDepDZ}

As already discussed, Ref.~\cite{DiCintio2014b} fitted a profile, similar to Equation (\ref{eq:rho_abc}),
to SPH simulations, finding the functional form of the three parameters in terms of $M_{\star}/M_{\rm vir}$, and Ref.~\cite{Lazar2020} proceeded 
similarly with the Einasto profile. The result provided a 
density profile with parameters depending on mass. 
\citet{Freundlich2020} did the same with the DZ profile. In summary, they 
fitted the logarithm of the density profile of the simulated haloes given by NIHAO, according to the DZ parameterization through a least-square minimization in the range \mbox{0.01 {$\le R_{\rm vir}\le 1$}}
. Given the mass $M_{\rm vir}$ of a 
profile, one can easily find $R_{\rm vir}$, 
and then one remains 
with two free parameters $a$, and $c$. One can proceed 
similarly with $M_{\star}$ and $M_{\star}/M_{\rm vir}$.
The profile radii $r$ were spaced logarithmically, with $N \simeq 100$ radii in the indicated range (\mbox{0.01 $R_{\rm vir}$--$R_{\rm vir}$}). As we already discussed, the parameters $s_1$ and $c_2$ are related to $a$ and $c$. The quality 
of the fit was evaluated using the rms of the residuals between the simulated $\log \rho$, and the model
\be
\sigma = \sqrt{\frac{1}{N}\sum_{i=1}^N \left(\log {\rho}_i - \log{\rho}_{\rm model}(r_i)\right)^2}
\ee

With the 
procedure described above, one can obtain the best fit $s_1$, and $c_2$, in terms of 
$M_{\rm vir}$, $M_{\star}$, and $M_{\star}/M_{\rm vir}$. In Figure 8 of~\citep{Freundlich2020}, the dependence of $s_1$ and $c_2$  is plotted 
in terms of $M_{\rm vir}$, $M_{\star}$, and $M_{\star}/M_{\rm vir}$, derived from the DZ density profile fits
.  

To capture the behavior of $s1$ and $c_2$ in terms of $M_{\rm vir}$, $M_{\star}$, and $M_{\star}/M_{\rm vir}$, \citet{Freundlich2020} assumed two functions
\be
\label{eq:s1(x)}
s_1(x) = \frac{s^\prime}{1+\left(\frac{x}{x_0}\right)^\nu} + s^{\prime\prime} \log \left(1+\left(\frac{x}{x_0}\right)^\nu\right),
\ee
where 
$x_0$, $s^\prime$, $s^{\prime\prime}$, and $\nu$ are adjustable parameters 
whose values are reproduced in Table 1 of~\citep{Freundlich2020}. 
Concerning $c_2$, they used 
\be
\label{eq:exp_function_c0}
c_2(x)= c^\prime \left(1+\left(\frac{x}{x_0} \right)^\nu\right)
\ee 
where again $x_0$, $c^\prime$, and $\nu$ are adjustable parameters.

With this procedure, the DZ becomes a mass-dependent profile, displaying 
some advantages over competing 
profiles. \textls[-15]{For example, 
compared with that of 
\citet{DiCintio2014b}}, it requires 
less free parameters and has analytic expressions for dispersion velocity, gravitational potential, and lensing properties. 

{Until }
now, we have described how~\citep{Freundlich2020} obtained a mass-dependent profile. In our case, we followed a similar procedure, 
with the difference that, while the simulated haloes' logarithm of the density profile was fitted by \citet{Freundlich2020} from the 
NIHAO simulations, according to 
the DZ parameterization, we used the results given by our model. The procedures are identical, except for the forms of 
the functions that 
capture the behavior of $s1$, and $c_2$ in terms of $M_{\rm vir}$, $M_{\star}$, and $M_{\star}/M_{\rm vir}$. From our model, we used the functions
\begin{eqnarray}
\label{eq:s1}
s_1(M_{\star}) \hspace{0.25cm} \textrm{or}\hspace{0.25cm}  c_2(M_{\star})&=& c_0+c_1 y +c_2 y^2+c_3 y^3+
c_4 y^4 \hspace{0.5cm} 10^4 \leq M_{\star} \leq 10^{12} \nonumber\\
y&=& \log10(1+(M_{\star}/x_0)^n),
\end{eqnarray}
while
\begin{eqnarray}
\label{eq:s11}
s_1(M_{\rm vir})&=& c_0+c_1 y+c_2 y^2+c_3 y^3+c_4 y^4+ 
+c_5 y^5 \hspace{0.5cm} 10^9 \leq M_{\rm vir} \leq 10^{15} \nonumber\\ 
y&=& \log10(1+(M_{\rm vir}/x_0)^n)
\end{eqnarray}
and
\begin{eqnarray}
\label{eq:c2}
c_2(M_{\rm vir})&=& c_0+c_1 y+c_2 y^2+c_3 y^3+ 
+c_4 y^4 \hspace{0.5cm} 10^{10.3} \leq M_{\rm vir} \leq 10^{15} \nonumber\\ 
y&=& \log10(1+(M_{\rm vir}/x_0)^n).
\end{eqnarray}

The parameters of the fit can be found in Table~\ref{table1}.

\begin{table}[H]

\caption{\label{table1}Parameters values for the fitting functions described in Equations~\eqref{eq:s1}--\eqref{eq:c2}
.}
\begin{adjustwidth}{-\extralength}{0cm}
\newcolumntype{C}{>{\centering\arraybackslash}X}

\begin{tabularx}{\fulllength}{p{1.5cm}ccccccccc}
\toprule
\textbf{Relation}  & \boldmath{$x_0$ [M$_{\odot}$] }&\boldmath{ $n$ }&\boldmath{ $c_0$}  &\boldmath{  $c_1$ }&\boldmath{ $c_2$} &\boldmath{ $c_3$ }&\boldmath{ $c_4$} &\boldmath{ $c_5$} &\boldmath{ $\sigma$}\\
\midrule
$s_1(M_{\star})$ & 2.25 $\times 10^7$ & 0.78 & 1.1  & $-$1.95 & 1.85 & $-$0.546 & 0.05 & - & 0.32\\
\midrule
$s_1(M_{\rm vir})$ & 2.1 $\times 10^{10}$ & 2.64 &0.93  & $-$0.8& 0.43& $-$0.0799 & 0.0063 & $-$1.84$\times 10^{-4}$ & 0.33\\
\midrule
$c_2(M_{\star})$ & 1.79 $\times 10^9$ &6.39 & 12.24 & 0.99867  & $-$9.99 $\times 10^{-5}$ &0.031 & $-$0.00198985  & -&  3.2\\
\midrule
$c_2(M_{\rm vir)}$ & 1.042 $\times 10^{14}$ & $-$1.88 & 6.94 & 15.93  & $-$3.39 & 0.0099005 & 0.0228 & - & 3.4\\
\bottomrule
\end{tabularx}
\end{adjustwidth}
\end{table}

\begin{figure}[H]
\begin{adjustwidth}{-\extralength}{0cm}
\centering
\includegraphics[width=0.98\columnwidth]{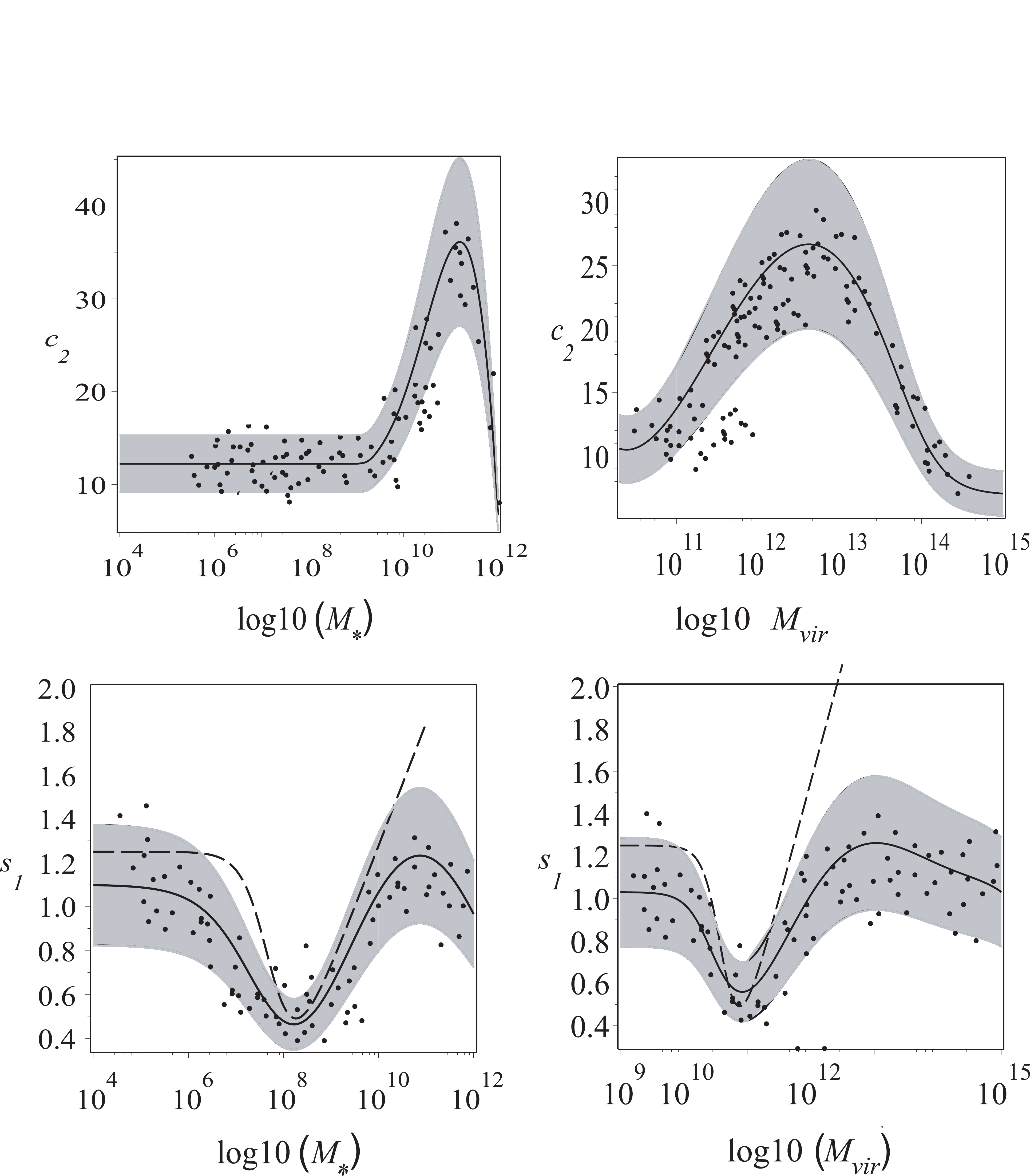}
\end{adjustwidth}
\caption{\label{fig:NSWD_branches_y01}
Mass dependence of the DZ 
 parameters $s_1$ and $c_2$ on 
$M_*$ and $M_{\rm vir}$. 
The parameters $s_1$ and $c_2$ are obtained from the fits to the density profile of $M_{\rm vir}$ and $M_{\star}$. The stellar-to-halo mass ratio $M_{\star}/M_{\rm vir}$ can be obtained from the other two profiles, as written in the text. 
The slope $s_1$ is obtained using 
Equations~(\ref{eq:s1}) and 
(\ref{eq:s11}). The best-fit curve is represented by the black solid line, while the dashed lines represent 
Equations~(45) and (46) of~\citep{Freundlich2020}. The parameters of the fit are given in Table~\ref{table1}. 
The parameter $c_2$ is obtained using 
Equations~(\ref{eq:s1}) and 
(\ref{eq:c2}).
The best-fit curve is represented by the black solid line. 
The rms $\sigma$ in gray is obtained, as discussed in {Section~\ref{sec:massDepDZ}}
, by means of an iterative process excluding points beyond 3 $\sigma$. }
\end{figure}

\vspace{-3pt} {}
Figure~\ref{fig:NSWD_branches_y01} shows the dependence of $s_1$ and $c_2$ on 
$M_{\star}$ and $M_{\rm vir}$. The shaded gray region 
represents the rms of the residuals. As described in Section~3.3.1 in \citep{Freundlich2020}, it is obtained by means of an iterative process excluding points beyond 3 $\sigma$. $s_1$ and $c_2$ are obtained from the fits to the density profile of $M_{\rm vir}$ and $M_{\star}$. The slope $s_1$ is obtained using 
\mbox{Equations~(\ref{eq:s1}) and (\ref{eq:s11})}. 
The black solid line represents the best-fit curve, while the dashed lines represent 
Equations~(45) and (46) of~\citep{Freundlich2020}. 
Similarly, parameter $c_2$ is obtained using Equations~(\ref{eq:s1}) and (\ref{eq:c2}). The best-fit curve is again represented by the black solid line. 
The 
relations of $s_1$ or 
$c_2$ in terms of $M_{\star}/M_{\rm vir}$ are 
not plotted, since it is directly related to the previous two relations (dependence on $M_{\rm vir}$ and $M_{\star}$). For a fixed value of $s_1$, one can obtain $M_{\star}$, and $M_{\rm vir}$, and consider their ratio. Moreover, the relation between $s_1$, $c_2$, and $M_{\star}/M_{\rm vir}$, because of 
AGN feedback, is characterized by a two-branch relation, as that shown in Figure 3 in \citep{Maccio2020}. In order to obtain the DZ DM profile relative to each halo, one can follow the prescription given in the following, also proposed in Section~4.1 in 
\citep{Freundlich2020}.
\begin{enumerate}[
label=(\arabic*),noitemsep,partopsep=0pt,topsep=0pt,parsep=0pt]
\item $\Mvir$ and the stellar mass $M_{\star}$ constitute the input parameters. These two quantities are related by abundance matching the $M_{\star}/M_{\rm vir}$ relation~\citep{Moster2013, Behroozi2013, Behroozi2019}.
\item The virial radius $\Rvir$ can be obtained using the relation 
\be
\Mvir = \frac{4\pi}{3} \Rvir^3 \Delta \rho_{\rm crit}
\ee
where $\Delta=18\pi^2 +82 x -39 x^2$, with $x=\Omega_m - 1$. As is well known, the critical density is given by $\rho_{\rm crit} = 3H^2/8\pi G$. Using \citet{Planck2014} parameters, $\Delta = 103.5$ and $\rho_{\rm crit}=124.9~\rm  M_\odot kpc^{-3}$. 
\item The parameters $s_1$ and $c_2$, in terms of $M_{\rm vir}$, $M_{\star}$, are given by 
Equations~(\ref{eq:s1})--(\ref{eq:s11})
with the parameters given in Table~\ref{table1}.
\item The parameters $a$ and $c$ of the DZ function can be obtained from the values of $s_1$ and $c_2$ using Equations~(\ref{eq:a(s1,c2)}) and (\ref{eq:c(s1,c2)}).
\item The scaling parameter $\rho_c$ is given by the following relations: $\mu = c^{a-3} (1+c^{1/2})^{2(3-a)}$, $\overline{\rho_{\rm vir}}=3\Mvir/4\pi \Rvir^3 =\Delta \rho_{\rm crit}$,  
and $\rho_c=(1-a/3) c^3 \mu \overline{\rho_{\rm vir}}$, where $r_c=\Rvir/c$.
\item Finally, the DZ mass-dependent density profile is given by Equation~(\ref{eq:rho32}). The corresponding 
circular velocity profile can be obtained from 
Equations~4 and 6 of~\citep{Freundlich2020} with 
$\rho_c=c^3\mu\overline{\rho_{\rm vir}}$, $b=2$, and $\overline{g}=3$.
\end{enumerate}
 

\vspace{-60pt} {}

An application of the method is shown in Figure~\ref{fig:einasto1}. The plot shows the density profiles for given values of the masses ($M_{\rm vir}$, $M_{\star}$, and $M_{\star}/M_{\rm vir}$) taken from the NIHAO suite. The red lines are the profiles from NIHAO, the short dashed lines those obtained by 
\citep{Freundlich2020} with the DZ mass-dependent profile, the dotted ones the result from \citet{DiCintio2014b}, and the long-dashed lines are the density profiles that we obtained. 
Similarly to   
\citep{Freundlich2020}, $s_1$ is obtained from Equations~(\ref{eq:s1})--(\ref{eq:s11}), while 
$c_2$, and $r_c$ are allowed to vary, and $\rho_c$ is constrained from the halo mass $M_{\rm vir}$. In the case of low masses, our density profiles are flatter at low radii. This is due to the fact that, in the NIHAO simulations, the role of the supernovae feedback is larger than in our model. As we discussed in Section~\ref{sec:Implementation}, in our case, the flattening of the profiles is mainly due to the interaction of gas clumps with DM through dynamical friction.

\section{Clusters of Galaxies' Density Profiles}\label{sec:clustersDen}

\begin{figure}[H]
\begin{adjustwidth}{-\extralength}{0cm}
\begin{centering}
\includegraphics[width=1.
2\columnwidth]{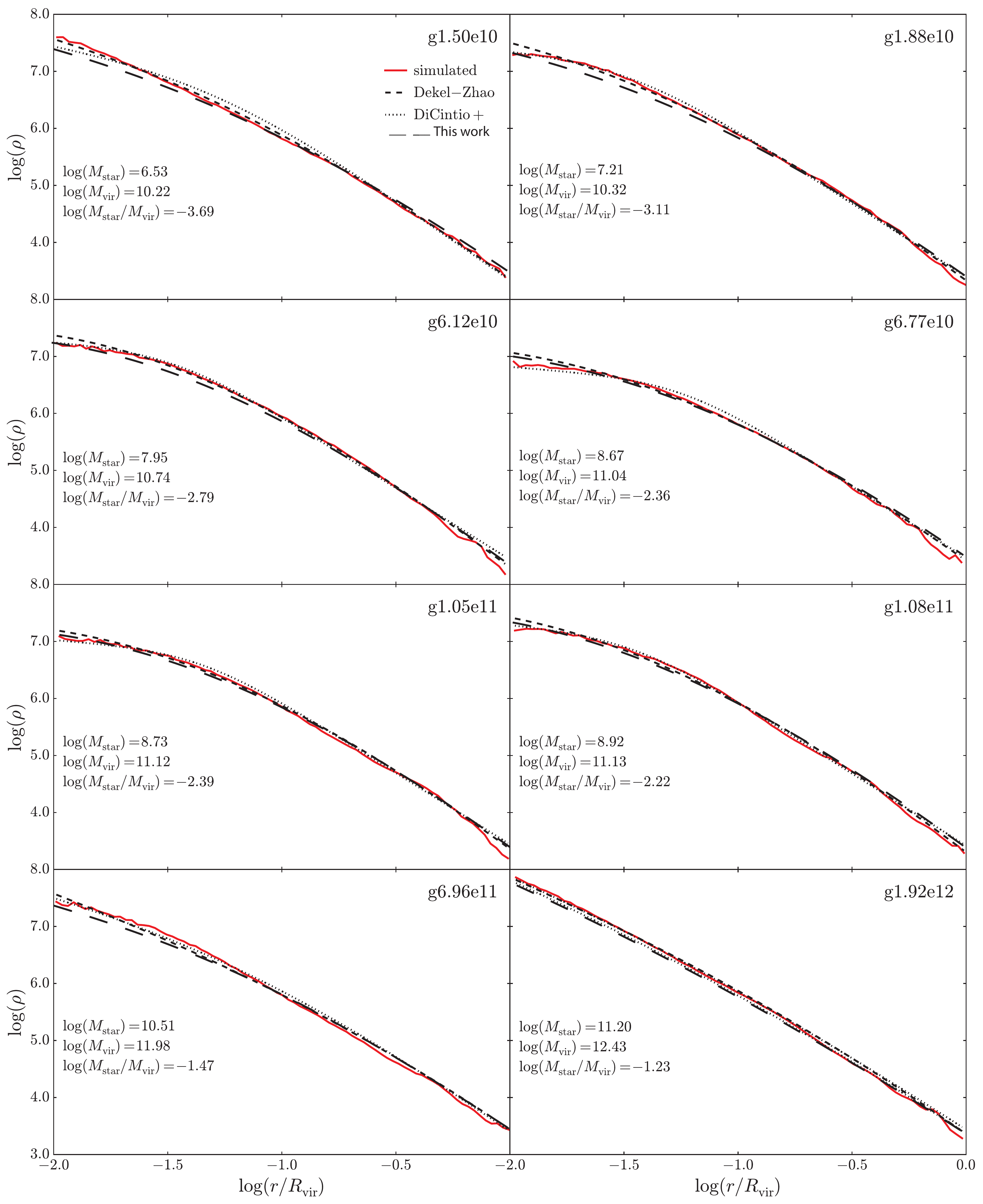}
\par\end{centering}
\end{adjustwidth}
\caption{\label{fig:einasto1} 
Comparison of the NIHAO 
 simulated density profile (red lines) with the DZ profile (short dashed line) of~\citep{Freundlich2020}, the \citet{DiCintio2014b} density profile 
(dotted line), and the DZ profile of our study (long dashed line). Similarly to   
\citep{Freundlich2020}, $s_1$ is obtained from Equations~(\ref{eq:s1})--(\ref{eq:s11}), 
$c_2$, an $r_c$ are allowed to vary, and $\rho_c$ is constrained from the halo mass 
$M_{\rm vir}$. $M_{\star} $, 
$M_{\rm vir}$ and the simulation galaxies' names 
are indicated. $M_{\star}/M_{\rm vir}$ is obtained by the ration of $M_{\star}$ and $M_{\rm vir}$. }
\end{figure}

\begin{figure}[H]
\begin{adjustwidth}{-\extralength}{0cm}
\begin{centering}
\includegraphics[width=1.2\columnwidth]{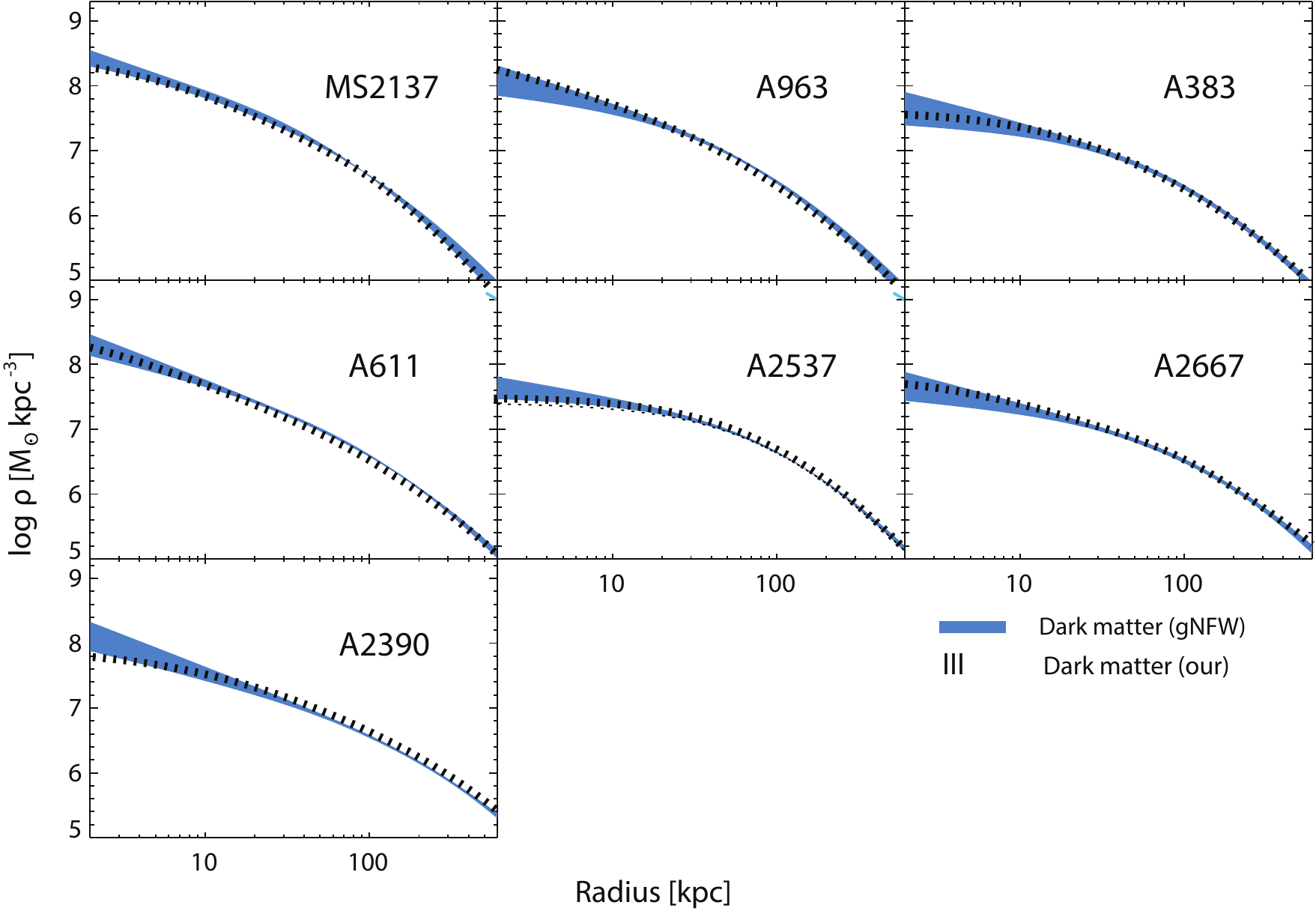}
\par\end{centering}
\end{adjustwidth}
\caption{\label{fig:einasto} Comparison of the 
 DM density profile obtained by~\citep{Newman2013a,Newman2013b} and the density profile predicted by the mass dependent DZ density profile. The width of the blue bands represents the 1 $\sigma$ uncertainty plus a systematic component (Section 4.3 of~\citep{Newman2013b}). The black bands represent the DZ mass dependent profile, and the 1 $\sigma$ uncertainty. }
\end{figure}

In Figure~\ref{fig:einasto1}, we showed the DZ density profile in the case of galaxy types 
from dwarfs to {the }Milky Way
. \citet{Freundlich2020} dedicated {their }
papers to show how the DZ mass dependent profile can be used to describe density profiles in better detail 
than the gNFW and Einasto, and furthermore compared it 
with~\citep{DiCintio2014b}. The main goal of the present paper is to extend 
\citep{Freundlich2020} to {the mass regime of galaxy }
clusters. To this aim, in Figure~\ref{fig:NSWD_branches_y01}, we obtained the $s_1$ and $c_2$ parameters in terms of $M_{\star}$, $M_{\rm vir}$, up to 
masses of the order of $10^{15} M_{\odot}$. In the following, we 
show how the DZ mass dependent profile can fit the density profile of clusters of galaxies. 
We choose the~\citep{Newman2013a,Newman2013b} clusters. As shown in Figure 3 of~\citep{Newman2013b}, the mass distribution was studied combining information from weak and strong lensing, and stellar kinematics, allowing 
the reconstruction of the mass distribution in baryons in the BCG. As shown in Figure~\ref{fig:einasto}, the total mass of the clusters studied (MS2137, A963, A383, A611, A2537, A2667, and A2390) is fitted by an NFW profile, while the DM distribution has profiles flatter than the NFW profile, as seen 
in Figure 5 of~\citep{Newman2013b}. The baryons usually dominate inside a 
radius of 10 kpc. Concerning the baryon content, Table 3 of~\citep{Newman2013a} gives the luminosity $L_V$ and Table 4 the ratio between stellar mass and luminosity, $\Upsilon=M_*/L_V$. The product of $\Upsilon$ by $L_V$ yields 
the stellar mass. Table 8, of the same paper, describes the characteristics of the total mass density profile, the virial mass, and radius, given 
$\Delta=200$, $M_{200}$ and 
$r_{200}$. {The }
results obtained by means of X-ray observations{ are also given}. We summarize in Table~\ref{table2} 
the characteristic of the clusters, starting with the inner slope calculated around 1~kpc, which corresponds to 
0.1 \% of the virial radius. We then 
report $L_V$, $\Upsilon$, the stellar mass, the virial mass, the ratio between stellar and virial mass, and finally 
the values of $s_1$ and $c_2$ obtained for each cluster with the method described in the previous sections. In Figure~\ref{fig:einasto}, we compared the DM density profile with DZ mass dependent profile calculated as described in the previous sections.  
The DM density profile obtained by~\citep{Newman2013a,Newman2013b} is represented in blue, while the mass dependent DZ prediction is plotted in black. The width of the blue bands represents the uncertainty, including the 1 $\sigma$ uncertainties for isotropic models, see~\citep{Newman2013a,Newman2013b}, and a systematic component obtained as described in Section~4.3 of \citep{Newman2013b}. 
As shown, there is a good agreement between the DM density profile and the mass dependent DZ profile.

\begin{table}[H]
\caption{\label{table2} Parameters of the clusters studied. Column 2 summarizes the inner DM slope, Column 3 the visual luminosity, Column 4 the mass-to-light ratio, Column 5 the baryon mass content, Column 6 the DM content, and Columns 7 and 8 the parameters $s_1$ and $c_2$.}
\newcolumntype{C}{>{\centering\arraybackslash}X}

\begin{tabularx}{\textwidth}{cccccccc}
\toprule
 & \boldmath{$\alpha$} &  \boldmath{$L_{v}\left(10^{11}\right)$} & \boldmath{ $\Upsilon$} & \boldmath{ $M_{\star}\left(10^{11}M_{<\odot}\right)$ }&  \boldmath{$\log_{10}\left(M_{200}/M_{\odot}\right)$ }&  \boldmath{$s_{1}$ }& \boldmath{ $c_{2}$}\tabularnewline
\midrule 
MS2137 & $0.65_{-0.30}^{+0.23}$ & $3.20$ & $2.05$ & $6.56$ & $14.56_{-0.18}^{+0.13}$ & $1.1$ & $6.01$\tabularnewline
\midrule 
A963 & $0.50_{-0.30}^{+0.27}$ & $4.61$ & $2.31$ & $10.65$ & $14.61_{-0.15}^{+0.11}$ & $1$ & $5.95$\tabularnewline
\midrule 
A383 & $0.37_{-0.23}^{+0.25}$ & $4.06$ & $2.26$ & $9.18$ & $14.82_{-0.08}^{+0.09}$ & $0.75$ & $5.9$\tabularnewline
\midrule 
A611 & $0.79_{-0.19}^{+0.14}$ & $5.47$ & $2.24$ & $12.25$ & $14.92{\scriptstyle \pm0.07}$ & $1.15$ & $5.86$\tabularnewline
\midrule 
A2537 & $0.23_{-0.16}^{+0.18}$ & $5.86$ & $2.32$ & $13.60$ & $15.12{\scriptstyle \pm0.04}$ & $0.72$ & $5.85$\tabularnewline
\midrule 
A2667 & $0.42_{-0.25}^{+0.23}$ & $3.89$ & $2.04$ & $7.94$ & $15.16{\scriptstyle \pm0.08}$ & $0.85$ & $5.57$\tabularnewline
\midrule 
A2390 & $0.82_{-0.18}^{+0.13}$ & $2.92$ & $1.80$ & $5.26$ & $15.34_{-0.07}^{+0.06}$ & $1.1$ & $4.45$\tabularnewline
\bottomrule 
\end{tabularx}

\end{table}

\section{Conclusions}\label{sec:conclusions}

In the present paper, we extended the work of \citet{Freundlich2020} related to the DZ density profile. By means of the NIHAO suite of simulations~\citep{Wang2015}, \citet{Dekel2017} showed that the functional form of Equation~(\ref{eq:rho_abc}) with $b=2$ and $g=3.5$ gives very good fits to haloes characterized by cores or 
cusps. This parameterization is dubbed DZ.
This function depends on two parameters: \citet{Freundlich2020} fitted the DZ profile to SPH simulations expressing the two parameters in terms of stellar mass, $M_*$, virial mass $M_v$, and their ratio. To capture the behavior of the two parameters, $s_1$ and 
$c_2$, in terms of the quoted masses, 
two peculiar functions for $s_1$, and $c_2${ were assumed}. In this way, the DZ profile {acquires a dependence on halo mass}
. While \citet{Freundlich2020} obtained a mass dependent DZ profile for the case of galaxies, we extended his work to clusters of galaxies.  
\citet{Freundlich2020} used the NIHAO simulations to obtain the mass dependence of the two parameters of the profile, while we used {our own analytical }
model
. In this model, baryonic clumps moving inside structures exchange angular momentum and energy with DM through dynamical friction. 
Following \citet{Freundlich2020}, we obtained the dependence from baryon physics of the two shape parameters. In this way, we obtained a mass dependent DZ profile describing the DM 
profiles from galaxies to clusters of galaxies. After obtaining the mass dependent DZ profile, we compared our model's 
simulated profiles 
with the results of~\citep{Freundlich2020,DiCintio2014b}. The results from 
our model are in good agreement with simulations and the corresponding 
profiles. As already reported, the main goal of the present paper is to extend 
\citep{Freundlich2020} up to 
the mass of clusters. To reach this goal, we obtained the $s_1$ and $c_2$ parameters in terms of $M_{\star}$, $M_{\rm vir}$, up to masses of the order of $10^{15} M_{\odot}$. In order to see how the DZ mass dependent profile fit the density profile of clusters of galaxies, we choose the~\citep{Newman2013a,Newman2013b} clusters. The mass distribution in these clusters was obtained combining several techniques, such as weak and strong lensing, and stellar kinematics. We studied the clusters MS2137, A963, A383, A611, A2537, A2667, and A2390. 
The DM distribution displays 
profiles flatter than the NFW profile, as shown in Figure 5 of~\citep{Newman2013b}. Given the parameters of the clusters obtained by~\citep{Newman2013a,Newman2013b}, we obtained values of $s_1$ and $c_2$ for each cluster with the method described in the previous sections. We compared the DM density profile obtained by~\citep{Newman2013a,Newman2013b} with those obtained with our model. 
As shown, there is a good agreement between the DM density profile and our 
mass dependent DZ profile.

%

%
%
\vspace{6pt} 



\authorcontributions{
All authors contributed equally to this work. All authors have read and agreed to the published version of the manuscript.}

\funding{MLeD acknowledges the financial support by 
the Lanzhou University starting
fund, the Fundamental Research Funds for the Central Universities
(Grant No. lzujbky-2019-25), the National Science Foundation of China (Grant No. 12047501), and the 111 Project under Grant No. B20063.}

\institutionalreview{Not applicable.}

\informedconsent{Not applicable.}

\dataavailability{
Not applicable. 
} 

\acknowledgments{The authors wish to thank Maksym Deliyergiyev for some calculations and Zhou Yong for the fits of Eqs~\eqref{eq:s1}-\eqref{eq:c2}.
}

\conflictsofinterest{The authors declare no conflict of interest.
} 

\begin{adjustwidth}{-\extralength}{0cm}

\reftitle{References}


\bibliography{old_MasterBib2,biblioNS,freu}

\begin{thebibliography}{999}

\bibitem[{Planck Collaboration} \em{et~al.}(2016){Planck Collaboration}, {Ade},
  {Aghanim}, {Arnaud}, {Ashdown}, {Aumont}, {Baccigalupi}, {Banday},
  {Barreiro}, {Bartlett}, and et~al.]{Planck2016}
{Planck Collaboration}.; {Ade}, P.A.R.; {Aghanim}, N.; {Arnaud}, M.; {Ashdown},
  M.; {Aumont}, J.; {Baccigalupi}, C.; {Banday}, A.J.; {Barreiro}, R.B.;
  {Bartlett}, J.G.;  et~al.
\newblock {Planck 2015 results. XIII. Cosmological parameters}.
\newblock {\em \aap} {\bf 2016}, {\em 594},~A13,
  \href{http://xxx.lanl.gov/abs/1502.01589}{{\normalfont [1502.01589]}}.
\newblock
  doi:{\changeurlcolor{black}\href{https://doi.org/10.1051/0004-6361/201525830}{\detokenize{10.1051/0004-6361/201525830}}}.

\bibitem[{Bertone} \em{et~al.}(2005){Bertone}, {Hooper}, and
  {Silk}]{Bertone2005}
{Bertone}, G.; {Hooper}, D.; {Silk}, J.
\newblock {Particle dark matter: evidence, candidates and constraints}.
\newblock {\em \physrep} {\bf 2005}, {\em 405},~279--390,
  \href{http://xxx.lanl.gov/abs/arXiv:hep-ph/0404175}{{\normalfont
  [arXiv:hep-ph/0404175]}}.
\newblock
  doi:{\changeurlcolor{black}\href{https://doi.org/10.1016/j.physrep.2004.08.031}{\detokenize{10.1016/j.physrep.2004.08.031}}}.

\bibitem[{Del Popolo}(2014)]{DelPopolo2014}
{Del Popolo}, A.
\newblock {Nonbaryonic Dark Matter in Cosmology}.
\newblock {\em International Journal of Modern Physics D} {\bf 2014}, {\em
  23},~30005,  \href{http://xxx.lanl.gov/abs/1305.0456}{{\normalfont
  [arXiv:astro-ph.CO/1305.0456]}}.
\newblock
  doi:{\changeurlcolor{black}\href{https://doi.org/10.1142/S0218271814300055}{\detokenize{10.1142/S0218271814300055}}}.

\bibitem[{Navarro} \em{et~al.}(1996){Navarro}, {Frenk}, and {White}]{NFW1996}
{Navarro}, J.F.; {Frenk}, C.S.; {White}, S.D.M.
\newblock {The Structure of Cold Dark Matter Halos}.
\newblock {\em \apj} {\bf 1996}, {\em 462},~563,
  \href{http://xxx.lanl.gov/abs/astro-ph/9508025}{{\normalfont
  [astro-ph/9508025]}}.
\newblock
  doi:{\changeurlcolor{black}\href{https://doi.org/10.1086/177173}{\detokenize{10.1086/177173}}}.

\bibitem[{Navarro} \em{et~al.}(1997){Navarro}, {Frenk}, and {White}]{NFW1997}
{Navarro}, J.F.; {Frenk}, C.S.; {White}, S.D.M.
\newblock {A Universal Density Profile from Hierarchical Clustering}.
\newblock {\em \apj} {\bf 1997}, {\em 490},~493--508,
  \href{http://xxx.lanl.gov/abs/astro-ph/9611107}{{\normalfont
  [astro-ph/9611107]}}.
\newblock
  doi:{\changeurlcolor{black}\href{https://doi.org/10.1086/304888}{\detokenize{10.1086/304888}}}.

\bibitem[{Navarro} \em{et~al.}(2004){Navarro}, {Hayashi}, {Power}, {Jenkins},
  {Frenk}, {White}, {Springel}, {Stadel}, and {Quinn}]{Navarro2004}
{Navarro}, J.F.; {Hayashi}, E.; {Power}, C.; {Jenkins}, A.R.; {Frenk}, C.S.;
  {White}, S.D.M.; {Springel}, V.; {Stadel}, J.; {Quinn}, T.R.
\newblock {The inner structure of {\ensuremath{\Lambda}}CDM haloes - III.
  Universality and asymptotic slopes}.
\newblock {\em \mnras} {\bf 2004}, {\em 349},~1039--1051,
  \href{http://xxx.lanl.gov/abs/astro-ph/0311231}{{\normalfont
  [arXiv:astro-ph/astro-ph/0311231]}}.
\newblock
  doi:{\changeurlcolor{black}\href{https://doi.org/10.1111/j.1365-2966.2004.07586.x}{\detokenize{10.1111/j.1365-2966.2004.07586.x}}}.

\bibitem[{Navarro} \em{et~al.}(2010){Navarro}, {Ludlow}, {Springel}, {Wang},
  {Vogelsberger}, {White}, {Jenkins}, {Frenk}, and {Helmi}]{Navarro2010}
{Navarro}, J.F.; {Ludlow}, A.; {Springel}, V.; {Wang}, J.; {Vogelsberger}, M.;
  {White}, S.D.M.; {Jenkins}, A.; {Frenk}, C.S.; {Helmi}, A.
\newblock {The diversity and similarity of simulated cold dark matter haloes}.
\newblock {\em \mnras} {\bf 2010}, {\em 402},~21--34,
  \href{http://xxx.lanl.gov/abs/0810.1522}{{\normalfont [0810.1522]}}.
\newblock
  doi:{\changeurlcolor{black}\href{https://doi.org/10.1111/j.1365-2966.2009.15878.x}{\detokenize{10.1111/j.1365-2966.2009.15878.x}}}.

\bibitem[{Gao} \em{et~al.}(2008){Gao}, {Navarro}, {Cole}, {Frenk}, {White},
  {Springel}, {Jenkins}, and {Neto}]{Gao2008}
{Gao}, L.; {Navarro}, J.F.; {Cole}, S.; {Frenk}, C.S.; {White}, S.D.M.;
  {Springel}, V.; {Jenkins}, A.; {Neto}, A.F.
\newblock {The redshift dependence of the structure of massive {$\Lambda$} cold
  dark matter haloes}.
\newblock {\em \mnras} {\bf 2008}, {\em 387},~536--544,
  \href{http://xxx.lanl.gov/abs/0711.0746}{{\normalfont [0711.0746]}}.
\newblock
  doi:{\changeurlcolor{black}\href{https://doi.org/10.1111/j.1365-2966.2008.13277.x}{\detokenize{10.1111/j.1365-2966.2008.13277.x}}}.

\bibitem[{Springel} \em{et~al.}(2008){Springel}, {Wang}, {Vogelsberger},
  {Ludlow}, {Jenkins}, {Helmi}, {Navarro}, {Frenk}, and {White}]{Springel2008}
{Springel}, V.; {Wang}, J.; {Vogelsberger}, M.; {Ludlow}, A.; {Jenkins}, A.;
  {Helmi}, A.; {Navarro}, J.F.; {Frenk}, C.S.; {White}, S.D.M.
\newblock {The Aquarius Project: the subhaloes of galactic haloes}.
\newblock {\em \mnras} {\bf 2008}, {\em 391},~1685--1711,
  \href{http://xxx.lanl.gov/abs/0809.0898}{{\normalfont
  [arXiv:astro-ph/0809.0898]}}.
\newblock
  doi:{\changeurlcolor{black}\href{https://doi.org/10.1111/j.1365-2966.2008.14066.x}{\detokenize{10.1111/j.1365-2966.2008.14066.x}}}.

\bibitem[{Flores} and {Primack}(1994)]{Flores1994}
{Flores}, R.A.; {Primack}, J.R.
\newblock {Observational and theoretical constraints on singular dark matter
  halos}.
\newblock {\em \apjl} {\bf 1994}, {\em 427},~L1--L4,
  \href{http://xxx.lanl.gov/abs/astro-ph/9402004}{{\normalfont
  [astro-ph/9402004]}}.
\newblock
  doi:{\changeurlcolor{black}\href{https://doi.org/10.1086/187350}{\detokenize{10.1086/187350}}}.

\bibitem[{Moore}(1994)]{Moore1994}
{Moore}, B.
\newblock {Evidence against dissipation-less dark matter from observations of
  galaxy haloes}.
\newblock {\em \nat} {\bf 1994}, {\em 370},~629--631.
\newblock
  doi:{\changeurlcolor{black}\href{https://doi.org/10.1038/370629a0}{\detokenize{10.1038/370629a0}}}.

\bibitem[{de Blok} \em{et~al.}(2008){de Blok}, {Walter}, {Brinks},
  {Trachternach}, {Oh}, and {Kennicutt}]{deBlok2008}
{de Blok}, W.J.G.; {Walter}, F.; {Brinks}, E.; {Trachternach}, C.; {Oh}, S.H.;
  {Kennicutt}, Jr., R.C.
\newblock {High-Resolution Rotation Curves and Galaxy Mass Models from THINGS}.
\newblock {\em \aj} {\bf 2008}, {\em 136},~2648--2719,
  \href{http://xxx.lanl.gov/abs/0810.2100}{{\normalfont [0810.2100]}}.
\newblock
  doi:{\changeurlcolor{black}\href{https://doi.org/10.1088/0004-6256/136/6/2648}{\detokenize{10.1088/0004-6256/136/6/2648}}}.

\bibitem[{de Blok}(2010)]{deBlok2010}
{de Blok}, W.J.G.
\newblock {The Core-Cusp Problem}.
\newblock {\em Advances in Astronomy} {\bf 2010}, {\em 2010},~789293,
  \href{http://xxx.lanl.gov/abs/0910.3538}{{\normalfont [0910.3538]}}.
\newblock
  doi:{\changeurlcolor{black}\href{https://doi.org/10.1155/2010/789293}{\detokenize{10.1155/2010/789293}}}.

\bibitem[{Kuzio de Naray} and {Spekkens}(2011)]{Kuzio2011}
{Kuzio de Naray}, R.; {Spekkens}, K.
\newblock {Do Baryons Alter the Halos of Low Surface Brightness Galaxies?}
\newblock {\em \apjl} {\bf 2011}, {\em 741},~L29,
  \href{http://xxx.lanl.gov/abs/1109.1288}{{\normalfont [1109.1288]}}.
\newblock
  doi:{\changeurlcolor{black}\href{https://doi.org/10.1088/2041-8205/741/2/L29}{\detokenize{10.1088/2041-8205/741/2/L29}}}.

\bibitem[{Oh} \em{et~al.}(2015){Oh}, {Hunter}, {Brinks}, {Elmegreen},
  {Schruba}, {Walter}, {Rupen}, {Young}, {Simpson}, {Johnson}, {Herrmann},
  {Ficut-Vicas}, {Cigan}, {Heesen}, {Ashley}, and {Zhang}]{Oh2015}
{Oh}, S.H.; {Hunter}, D.A.; {Brinks}, E.; {Elmegreen}, B.G.; {Schruba}, A.;
  {Walter}, F.; {Rupen}, M.P.; {Young}, L.M.; {Simpson}, C.E.; {Johnson}, M.C.;
   et~al.
\newblock {High-resolution Mass Models of Dwarf Galaxies from LITTLE THINGS}.
\newblock {\em \aj} {\bf 2015}, {\em 149},~180,
  \href{http://xxx.lanl.gov/abs/1502.01281}{{\normalfont [1502.01281]}}.
\newblock
  doi:{\changeurlcolor{black}\href{https://doi.org/10.1088/0004-6256/149/6/180}{\detokenize{10.1088/0004-6256/149/6/180}}}.

\bibitem[{Newman} \em{et~al.}(2013{\natexlab{a}}){Newman}, {Treu}, {Ellis},
  {Sand}, {Nipoti}, {Richard}, and {Jullo}]{Newman2013a}
{Newman}, A.B.; {Treu}, T.; {Ellis}, R.S.; {Sand}, D.J.; {Nipoti}, C.;
  {Richard}, J.; {Jullo}, E.
\newblock {The Density Profiles of Massive, Relaxed Galaxy Clusters. I. The
  Total Density Over Three Decades in Radius}.
\newblock {\em \apj} {\bf 2013}, {\em 765},~24,
  \href{http://xxx.lanl.gov/abs/1209.1391}{{\normalfont
  [arXiv:astro-ph.CO/1209.1391]}}.
\newblock
  doi:{\changeurlcolor{black}\href{https://doi.org/10.1088/0004-637X/765/1/24}{\detokenize{10.1088/0004-637X/765/1/24}}}.

\bibitem[{Newman} \em{et~al.}(2013{\natexlab{b}}){Newman}, {Treu}, {Ellis}, and
  {Sand}]{Newman2013b}
{Newman}, A.B.; {Treu}, T.; {Ellis}, R.S.; {Sand}, D.J.
\newblock {The Density Profiles of Massive, Relaxed Galaxy Clusters. II.
  Separating Luminous and Dark Matter in Cluster Cores}.
\newblock {\em \apj} {\bf 2013}, {\em 765},~25,
  \href{http://xxx.lanl.gov/abs/1209.1392}{{\normalfont
  [arXiv:astro-ph.CO/1209.1392]}}.
\newblock
  doi:{\changeurlcolor{black}\href{https://doi.org/10.1088/0004-637X/765/1/25}{\detokenize{10.1088/0004-637X/765/1/25}}}.

\bibitem[{Adams} \em{et~al.}(2014){Adams}, {Simon}, {Fabricius}, {van den
  Bosch}, {Barentine}, {Bender}, {Gebhardt}, {Hill}, {Murphy}, {Swaters},
  {Thomas}, and {van de Ven}]{Adams2014}
{Adams}, J.J.; {Simon}, J.D.; {Fabricius}, M.H.; {van den Bosch}, R.C.E.;
  {Barentine}, J.C.; {Bender}, R.; {Gebhardt}, K.; {Hill}, G.J.; {Murphy},
  J.D.; {Swaters}, R.A.;  et~al.
\newblock {Dwarf Galaxy Dark Matter Density Profiles Inferred from Stellar and
  Gas Kinematics}.
\newblock {\em \apj} {\bf 2014}, {\em 789},~63,
  \href{http://xxx.lanl.gov/abs/1405.4854}{{\normalfont [1405.4854]}}.
\newblock
  doi:{\changeurlcolor{black}\href{https://doi.org/10.1088/0004-637X/789/1/63}{\detokenize{10.1088/0004-637X/789/1/63}}}.

\bibitem[{Burkert}(1995)]{Burkert1995}
{Burkert}, A.
\newblock {The Structure of Dark Matter Halos in Dwarf Galaxies}.
\newblock {\em \apjl} {\bf 1995}, {\em 447},~L25,
  \href{http://xxx.lanl.gov/abs/astro-ph/9504041}{{\normalfont
  [astro-ph/9504041]}}.
\newblock
  doi:{\changeurlcolor{black}\href{https://doi.org/10.1086/309560}{\detokenize{10.1086/309560}}}.

\bibitem[{de Blok} \em{et~al.}(2003){de Blok}, {Bosma}, and
  {McGaugh}]{deBlok2003}
{de Blok}, W.J.G.; {Bosma}, A.; {McGaugh}, S.
\newblock {Simulating observations of dark matter dominated galaxies: towards
  the optimal halo profile}.
\newblock {\em \mnras} {\bf 2003}, {\em 340},~657--678,
  \href{http://xxx.lanl.gov/abs/astro-ph/0212102}{{\normalfont
  [astro-ph/0212102]}}.
\newblock
  doi:{\changeurlcolor{black}\href{https://doi.org/10.1046/j.1365-8711.2003.06330.x}{\detokenize{10.1046/j.1365-8711.2003.06330.x}}}.

\bibitem[{Swaters} \em{et~al.}(2003){Swaters}, {Madore}, {van den Bosch}, and
  {Balcells}]{Swaters2003}
{Swaters}, R.A.; {Madore}, B.F.; {van den Bosch}, F.C.; {Balcells}, M.
\newblock {The Central Mass Distribution in Dwarf and Low Surface Brightness
  Galaxies}.
\newblock {\em \apj} {\bf 2003}, {\em 583},~732--751,
  \href{http://xxx.lanl.gov/abs/astro-ph/0210152}{{\normalfont
  [astro-ph/0210152]}}.
\newblock
  doi:{\changeurlcolor{black}\href{https://doi.org/10.1086/345426}{\detokenize{10.1086/345426}}}.

\bibitem[{Kuzio de Naray} and {Kaufmann}(2011)]{KuziodeNaray2011}
{Kuzio de Naray}, R.; {Kaufmann}, T.
\newblock {Recovering cores and cusps in dark matter haloes using mock velocity
  field observations}.
\newblock {\em \mnras} {\bf 2011}, {\em 414},~3617--3626,
  \href{http://xxx.lanl.gov/abs/1012.3471}{{\normalfont
  [arXiv:astro-ph.CO/1012.3471]}}.
\newblock
  doi:{\changeurlcolor{black}\href{https://doi.org/10.1111/j.1365-2966.2011.18656.x}{\detokenize{10.1111/j.1365-2966.2011.18656.x}}}.

\bibitem[{Oh} \em{et~al.}(2011{\natexlab{a}}){Oh}, {de Blok}, {Brinks},
  {Walter}, and {Kennicutt}]{Oh2011a}
{Oh}, S.H.; {de Blok}, W.J.G.; {Brinks}, E.; {Walter}, F.; {Kennicutt}, Jr.,
  R.C.
\newblock {Dark and Luminous Matter in THINGS Dwarf Galaxies}.
\newblock {\em \aj} {\bf 2011}, {\em 141},~193,
  \href{http://xxx.lanl.gov/abs/1011.0899}{{\normalfont
  [arXiv:astro-ph.CO/1011.0899]}}.
\newblock
  doi:{\changeurlcolor{black}\href{https://doi.org/10.1088/0004-6256/141/6/193}{\detokenize{10.1088/0004-6256/141/6/193}}}.

\bibitem[{Oh} \em{et~al.}(2011{\natexlab{b}}){Oh}, {Brook}, {Governato},
  {Brinks}, {Mayer}, {de Blok}, {Brooks}, and {Walter}]{Oh2011b}
{Oh}, S.H.; {Brook}, C.; {Governato}, F.; {Brinks}, E.; {Mayer}, L.; {de Blok},
  W.J.G.; {Brooks}, A.; {Walter}, F.
\newblock {The Central Slope of Dark Matter Cores in Dwarf Galaxies:
  Simulations versus THINGS}.
\newblock {\em \aj} {\bf 2011}, {\em 142},~24,
  \href{http://xxx.lanl.gov/abs/1011.2777}{{\normalfont
  [arXiv:astro-ph.CO/1011.2777]}}.
\newblock
  doi:{\changeurlcolor{black}\href{https://doi.org/10.1088/0004-6256/142/1/24}{\detokenize{10.1088/0004-6256/142/1/24}}}.

\bibitem[{Governato} \em{et~al.}(2010){Governato}, {Brook}, {Mayer}, {Brooks},
  {Rhee}, {Wadsley}, {Jonsson}, {Willman}, {Stinson}, {Quinn}, and
  {Madau}]{Governato2010}
{Governato}, F.; {Brook}, C.; {Mayer}, L.; {Brooks}, A.; {Rhee}, G.; {Wadsley},
  J.; {Jonsson}, P.; {Willman}, B.; {Stinson}, G.; {Quinn}, T.;  et~al.
\newblock {Bulgeless dwarf galaxies and dark matter cores from supernova-driven
  outflows}.
\newblock {\em \nat} {\bf 2010}, {\em 463},~203--206,
  \href{http://xxx.lanl.gov/abs/0911.2237}{{\normalfont
  [arXiv:astro-ph.CO/0911.2237]}}.
\newblock
  doi:{\changeurlcolor{black}\href{https://doi.org/10.1038/nature08640}{\detokenize{10.1038/nature08640}}}.

\bibitem[{Governato} \em{et~al.}(2012){Governato}, {Zolotov}, {Pontzen},
  {Christensen}, {Oh}, {Brooks}, {Quinn}, {Shen}, and {Wadsley}]{Governato2012}
{Governato}, F.; {Zolotov}, A.; {Pontzen}, A.; {Christensen}, C.; {Oh}, S.H.;
  {Brooks}, A.M.; {Quinn}, T.; {Shen}, S.; {Wadsley}, J.
\newblock {Cuspy no more: how outflows affect the central dark matter and
  baryon distribution in {$\Lambda$} cold dark matter galaxies}.
\newblock {\em \mnras} {\bf 2012}, {\em 422},~1231--1240,
  \href{http://xxx.lanl.gov/abs/1202.0554}{{\normalfont
  [arXiv:astro-ph.CO/1202.0554]}}.
\newblock
  doi:{\changeurlcolor{black}\href{https://doi.org/10.1111/j.1365-2966.2012.20696.x}{\detokenize{10.1111/j.1365-2966.2012.20696.x}}}.

\bibitem[{Del Popolo}(2009)]{DelPopolo2009}
{Del Popolo}, A.
\newblock {The Cusp/Core Problem and the Secondary Infall Model}.
\newblock {\em \apj} {\bf 2009}, {\em 698},~2093--2113,
  \href{http://xxx.lanl.gov/abs/0906.4447}{{\normalfont
  [arXiv:astro-ph.CO/0906.4447]}}.
\newblock
  doi:{\changeurlcolor{black}\href{https://doi.org/10.1088/0004-637X/698/2/2093}{\detokenize{10.1088/0004-637X/698/2/2093}}}.

\bibitem[{Cardone} and {Del Popolo}(2012)]{Cardone2012}
{Cardone}, V.F.; {Del Popolo}, A.
\newblock {Newtonian acceleration scales in spiral galaxies}.
\newblock {\em \mnras} {\bf 2012}, {\em 427},~3176--3187,
  \href{http://xxx.lanl.gov/abs/1209.1524}{{\normalfont
  [arXiv:astro-ph.CO/1209.1524]}}.
\newblock
  doi:{\changeurlcolor{black}\href{https://doi.org/10.1111/j.1365-2966.2012.21982.x}{\detokenize{10.1111/j.1365-2966.2012.21982.x}}}.

\bibitem[{Del Popolo}(2012{\natexlab{a}})]{DelPopolo2012a}
{Del Popolo}, A.
\newblock {Density profile slopes of dwarf galaxies and their environment}.
\newblock {\em \mnras} {\bf 2012}, {\em 419},~971--984,
  \href{http://xxx.lanl.gov/abs/1105.0090}{{\normalfont
  [arXiv:astro-ph.CO/1105.0090]}}.
\newblock
  doi:{\changeurlcolor{black}\href{https://doi.org/10.1111/j.1365-2966.2011.19754.x}{\detokenize{10.1111/j.1365-2966.2011.19754.x}}}.

\bibitem[{Del Popolo}(2012{\natexlab{b}})]{DelPopolo2012b}
{Del Popolo}, A.
\newblock {On the density-profile slope of clusters of galaxies}.
\newblock {\em \mnras} {\bf 2012}, {\em 424},~38--51,
  \href{http://xxx.lanl.gov/abs/1204.4439}{{\normalfont
  [arXiv:astro-ph.CO/1204.4439]}}.
\newblock
  doi:{\changeurlcolor{black}\href{https://doi.org/10.1111/j.1365-2966.2012.21141.x}{\detokenize{10.1111/j.1365-2966.2012.21141.x}}}.

\bibitem[{Del Popolo} and {Hiotelis}(2014)]{DelPopolo2014a}
{Del Popolo}, A.; {Hiotelis}, N.
\newblock {Cusps and cores in the presence of galactic bulges}.
\newblock {\em \jcap} {\bf 2014}, {\em 1},~47,
  \href{http://xxx.lanl.gov/abs/1401.6577}{{\normalfont
  [arXiv:astro-ph.GA/1401.6577]}}.
\newblock
  doi:{\changeurlcolor{black}\href{https://doi.org/10.1088/1475-7516/2014/01/047}{\detokenize{10.1088/1475-7516/2014/01/047}}}.

\bibitem[{Moore} \em{et~al.}(1999){Moore}, {Quinn}, {Governato}, {Stadel}, and
  {Lake}]{Moore1999}
{Moore}, B.; {Quinn}, T.; {Governato}, F.; {Stadel}, J.; {Lake}, G.
\newblock {Cold collapse and the core catastrophe}.
\newblock {\em \mnras} {\bf 1999}, {\em 310},~1147--1152,
  \href{http://xxx.lanl.gov/abs/astro-ph/9903164}{{\normalfont
  [astro-ph/9903164]}}.
\newblock
  doi:{\changeurlcolor{black}\href{https://doi.org/10.1046/j.1365-8711.1999.03039.x}{\detokenize{10.1046/j.1365-8711.1999.03039.x}}}.

\bibitem[{Boylan-Kolchin} \em{et~al.}(2011){Boylan-Kolchin}, {Bullock}, and
  {Kaplinghat}]{BoylanKolchin2011}
{Boylan-Kolchin}, M.; {Bullock}, J.S.; {Kaplinghat}, M.
\newblock {Too big to fail? The puzzling darkness of massive Milky Way
  subhaloes}.
\newblock {\em \mnras} {\bf 2011}, {\em 415},~L40--L44,
  \href{http://xxx.lanl.gov/abs/1103.0007}{{\normalfont
  [arXiv:astro-ph.CO/1103.0007]}}.
\newblock
  doi:{\changeurlcolor{black}\href{https://doi.org/10.1111/j.1745-3933.2011.01074.x}{\detokenize{10.1111/j.1745-3933.2011.01074.x}}}.

\bibitem[{Boylan-Kolchin} \em{et~al.}(2012){Boylan-Kolchin}, {Bullock}, and
  {Kaplinghat}]{BoylanKolchin2012}
{Boylan-Kolchin}, M.; {Bullock}, J.S.; {Kaplinghat}, M.
\newblock {The Milky Way's bright satellites as an apparent failure of
  {$\Lambda$}CDM}.
\newblock {\em \mnras} {\bf 2012}, {\em 422},~1203--1218,
  \href{http://xxx.lanl.gov/abs/1111.2048}{{\normalfont
  [arXiv:astro-ph.CO/1111.2048]}}.
\newblock
  doi:{\changeurlcolor{black}\href{https://doi.org/10.1111/j.1365-2966.2012.20695.x}{\detokenize{10.1111/j.1365-2966.2012.20695.x}}}.

\bibitem[{Sand} \em{et~al.}(2002){Sand}, {Treu}, and {Ellis}]{Sand2002}
{Sand}, D.J.; {Treu}, T.; {Ellis}, R.S.
\newblock {The Dark Matter Density Profile of the Lensing Cluster MS 2137-23: A
  Test of the Cold Dark Matter Paradigm}.
\newblock {\em \apjl} {\bf 2002}, {\em 574},~L129--L133,
  \href{http://xxx.lanl.gov/abs/astro-ph/0207048}{{\normalfont
  [astro-ph/0207048]}}.
\newblock
  doi:{\changeurlcolor{black}\href{https://doi.org/10.1086/342530}{\detokenize{10.1086/342530}}}.

\bibitem[{Sand} \em{et~al.}(2004){Sand}, {Treu}, {Smith}, and
  {Ellis}]{Sand2004}
{Sand}, D.J.; {Treu}, T.; {Smith}, G.P.; {Ellis}, R.S.
\newblock {The Dark Matter Distribution in the Central Regions of Galaxy
  Clusters: Implications for Cold Dark Matter}.
\newblock {\em \apj} {\bf 2004}, {\em 604},~88--107,
  \href{http://xxx.lanl.gov/abs/astro-ph/0309465}{{\normalfont
  [astro-ph/0309465]}}.
\newblock
  doi:{\changeurlcolor{black}\href{https://doi.org/10.1086/382146}{\detokenize{10.1086/382146}}}.

\bibitem[{Col{\'{\i}}n} \em{et~al.}(2000){Col{\'{\i}}n}, {Avila-Reese}, and
  {Valenzuela}]{Colin2000}
{Col{\'{\i}}n}, P.; {Avila-Reese}, V.; {Valenzuela}, O.
\newblock {Substructure and Halo Density Profiles in a Warm Dark Matter
  Cosmology}.
\newblock {\em \apj} {\bf 2000}, {\em 542},~622--630,
  \href{http://xxx.lanl.gov/abs/astro-ph/0004115}{{\normalfont
  [astro-ph/0004115]}}.
\newblock
  doi:{\changeurlcolor{black}\href{https://doi.org/10.1086/317057}{\detokenize{10.1086/317057}}}.

\bibitem[{Goodman}(2000)]{Goodman2000}
{Goodman}, J.
\newblock {Repulsive dark matter}.
\newblock {\em \na} {\bf 2000}, {\em 5},~103--107,
  \href{http://xxx.lanl.gov/abs/astro-ph/0003018}{{\normalfont
  [astro-ph/0003018]}}.
\newblock
  doi:{\changeurlcolor{black}\href{https://doi.org/10.1016/S1384-1076(00)00015-4}{\detokenize{10.1016/S1384-1076(00)00015-4}}}.

\bibitem[{Hu} \em{et~al.}(2000){Hu}, {Barkana}, and {Gruzinov}]{Hu2000}
{Hu}, W.; {Barkana}, R.; {Gruzinov}, A.
\newblock {Fuzzy Cold Dark Matter: The Wave Properties of Ultralight
  Particles}.
\newblock {\em Physical Review Letters} {\bf 2000}, {\em 85},~1158,
  \href{http://xxx.lanl.gov/abs/astro-ph/0003365}{{\normalfont
  [astro-ph/0003365]}}.
\newblock
  doi:{\changeurlcolor{black}\href{https://doi.org/10.1103/PhysRevLett.85.1158}{\detokenize{10.1103/PhysRevLett.85.1158}}}.

\bibitem[{Kaplinghat} \em{et~al.}(2000){Kaplinghat}, {Knox}, and
  {Turner}]{Kaplinghat2000}
{Kaplinghat}, M.; {Knox}, L.; {Turner}, M.S.
\newblock {Annihilating Cold Dark Matter}.
\newblock {\em Physical Review Letters} {\bf 2000}, {\em 85},~3335,
  \href{http://xxx.lanl.gov/abs/astro-ph/0005210}{{\normalfont
  [astro-ph/0005210]}}.
\newblock
  doi:{\changeurlcolor{black}\href{https://doi.org/10.1103/PhysRevLett.85.3335}{\detokenize{10.1103/PhysRevLett.85.3335}}}.

\bibitem[{Peebles}(2000)]{Peebles2000}
{Peebles}, P.J.E.
\newblock {Fluid Dark Matter}.
\newblock {\em \apjl} {\bf 2000}, {\em 534},~L127--L129,
  \href{http://xxx.lanl.gov/abs/astro-ph/0002495}{{\normalfont
  [astro-ph/0002495]}}.
\newblock
  doi:{\changeurlcolor{black}\href{https://doi.org/10.1086/312677}{\detokenize{10.1086/312677}}}.

\bibitem[{Sommer-Larsen} and {Dolgov}(2001)]{SommerLarsen2001}
{Sommer-Larsen}, J.; {Dolgov}, A.
\newblock {Formation of Disk Galaxies: Warm Dark Matter and the Angular
  Momentum Problem}.
\newblock {\em \apj} {\bf 2001}, {\em 551},~608--623,
  \href{http://xxx.lanl.gov/abs/astro-ph/9912166}{{\normalfont
  [astro-ph/9912166]}}.
\newblock
  doi:{\changeurlcolor{black}\href{https://doi.org/10.1086/320211}{\detokenize{10.1086/320211}}}.

\bibitem[{Zentner} and {Bullock}(2003)]{Zentner2003}
{Zentner}, A.R.; {Bullock}, J.S.
\newblock {Halo Substructure and the Power Spectrum}.
\newblock {\em \apj} {\bf 2003}, {\em 598},~49--72,
  \href{http://xxx.lanl.gov/abs/astro-ph/0304292}{{\normalfont
  [astro-ph/0304292]}}.
\newblock
  doi:{\changeurlcolor{black}\href{https://doi.org/10.1086/378797}{\detokenize{10.1086/378797}}}.

\bibitem[{Buchdahl}(1970)]{Buchdahl1970}
{Buchdahl}, H.A.
\newblock {Non-linear Lagrangians and cosmological theory}.
\newblock {\em \mnras} {\bf 1970}, {\em 150},~1.

\bibitem[{Starobinsky}(1980)]{Starobinsky1980}
{Starobinsky}, A.A.
\newblock {A new type of isotropic cosmological models without singularity}.
\newblock {\em Physics Letters B} {\bf 1980}, {\em 91},~99--102.
\newblock
  doi:{\changeurlcolor{black}\href{https://doi.org/10.1016/0370-2693(80)90670-X}{\detokenize{10.1016/0370-2693(80)90670-X}}}.

\bibitem[{Bengochea} and {Ferraro}(2009)]{Bengochea2009}
{Bengochea}, G.R.; {Ferraro}, R.
\newblock {Dark torsion as the cosmic speed-up}.
\newblock {\em \prd} {\bf 2009}, {\em 79},~124019--+,
  \href{http://xxx.lanl.gov/abs/0812.1205}{{\normalfont [0812.1205]}}.
\newblock
  doi:{\changeurlcolor{black}\href{https://doi.org/10.1103/PhysRevD.79.124019}{\detokenize{10.1103/PhysRevD.79.124019}}}.

\bibitem[{Linder}(2010)]{Linder2010}
{Linder}, E.V.
\newblock {Einstein's other gravity and the acceleration of the Universe}.
\newblock {\em \prd} {\bf 2010}, {\em 81},~127301--+,
  \href{http://xxx.lanl.gov/abs/1005.3039}{{\normalfont
  [arXiv:astro-ph.CO/1005.3039]}}.
\newblock
  doi:{\changeurlcolor{black}\href{https://doi.org/10.1103/PhysRevD.81.127301}{\detokenize{10.1103/PhysRevD.81.127301}}}.

\bibitem[{Dent} \em{et~al.}(2011){Dent}, {Dutta}, and {Saridakis}]{Dent2011}
{Dent}, J.B.; {Dutta}, S.; {Saridakis}, E.N.
\newblock {f(T) gravity mimicking dynamical dark energy. Background and
  perturbation analysis}.
\newblock {\em \jcap} {\bf 2011}, {\em 1},~9--+,
  \href{http://xxx.lanl.gov/abs/1010.2215}{{\normalfont
  [arXiv:astro-ph.CO/1010.2215]}}.
\newblock
  doi:{\changeurlcolor{black}\href{https://doi.org/10.1088/1475-7516/2011/01/009}{\detokenize{10.1088/1475-7516/2011/01/009}}}.

\bibitem[{Zheng} and {Huang}(2011)]{Zheng2011}
{Zheng}, R.; {Huang}, Q.G.
\newblock {Growth factor in f(T) gravity}.
\newblock {\em \jcap} {\bf 2011}, {\em 3},~2--+,
  \href{http://xxx.lanl.gov/abs/1010.3512}{{\normalfont
  [arXiv:gr-qc/1010.3512]}}.
\newblock
  doi:{\changeurlcolor{black}\href{https://doi.org/10.1088/1475-7516/2011/03/002}{\detokenize{10.1088/1475-7516/2011/03/002}}}.

\bibitem[{Milgrom}(1983{\natexlab{a}})]{Milgrom1983b}
{Milgrom}, M.
\newblock {A modification of the Newtonian dynamics - Implications for
  galaxies}.
\newblock {\em \apj} {\bf 1983}, {\em 270},~371--389.
\newblock
  doi:{\changeurlcolor{black}\href{https://doi.org/10.1086/161131}{\detokenize{10.1086/161131}}}.

\bibitem[{Milgrom}(1983{\natexlab{b}})]{Milgrom1983a}
{Milgrom}, M.
\newblock {A modification of the Newtonian dynamics as a possible alternative
  to the hidden mass hypothesis}.
\newblock {\em \apj} {\bf 1983}, {\em 270},~365--370.
\newblock
  doi:{\changeurlcolor{black}\href{https://doi.org/10.1086/161130}{\detokenize{10.1086/161130}}}.

\bibitem[{Navarro} \em{et~al.}(1996){Navarro}, {Eke}, and
  {Frenk}]{Navarro1996a}
{Navarro}, J.F.; {Eke}, V.R.; {Frenk}, C.S.
\newblock {The cores of dwarf galaxy haloes}.
\newblock {\em \mnras} {\bf 1996}, {\em 283},~L72--L78,
  \href{http://xxx.lanl.gov/abs/astro-ph/9610187}{{\normalfont
  [astro-ph/9610187]}}.

\bibitem[{Gelato} and {Sommer-Larsen}(1999)]{Gelato1999}
{Gelato}, S.; {Sommer-Larsen}, J.
\newblock {On DDO 154 and cold dark matter halo profiles}.
\newblock {\em \mnras} {\bf 1999}, {\em 303},~321--328,
  \href{http://xxx.lanl.gov/abs/astro-ph/9806289}{{\normalfont
  [astro-ph/9806289]}}.
\newblock
  doi:{\changeurlcolor{black}\href{https://doi.org/10.1046/j.1365-8711.1999.02223.x}{\detokenize{10.1046/j.1365-8711.1999.02223.x}}}.

\bibitem[{Read} and {Gilmore}(2005)]{Read2005}
{Read}, J.I.; {Gilmore}, G.
\newblock {Mass loss from dwarf spheroidal galaxies: the origins of shallow
  dark matter cores and exponential surface brightness profiles}.
\newblock {\em \mnras} {\bf 2005}, {\em 356},~107--124,
  \href{http://xxx.lanl.gov/abs/astro-ph/0409565}{{\normalfont
  [astro-ph/0409565]}}.
\newblock
  doi:{\changeurlcolor{black}\href{https://doi.org/10.1111/j.1365-2966.2004.08424.x}{\detokenize{10.1111/j.1365-2966.2004.08424.x}}}.

\bibitem[{Mashchenko} \em{et~al.}(2006){Mashchenko}, {Couchman}, and
  {Sills}]{Mashchenko2006}
{Mashchenko}, S.; {Couchman}, H.M.P.; {Sills}, A.
\newblock {Formation of Minigalaxies in Defunct Cosmological H II Regions}.
\newblock {\em \apj} {\bf 2006}, {\em 639},~633--643,
  \href{http://xxx.lanl.gov/abs/astro-ph/0511361}{{\normalfont
  [astro-ph/0511361]}}.
\newblock
  doi:{\changeurlcolor{black}\href{https://doi.org/10.1086/499582}{\detokenize{10.1086/499582}}}.

\bibitem[{Mashchenko} \em{et~al.}(2008){Mashchenko}, {Wadsley}, and
  {Couchman}]{Mashchenko2008}
{Mashchenko}, S.; {Wadsley}, J.; {Couchman}, H.M.P.
\newblock {Stellar Feedback in Dwarf Galaxy Formation}.
\newblock {\em Science} {\bf 2008}, {\em 319},~174--,
  \href{http://xxx.lanl.gov/abs/0711.4803}{{\normalfont [0711.4803]}}.
\newblock
  doi:{\changeurlcolor{black}\href{https://doi.org/10.1126/science.1148666}{\detokenize{10.1126/science.1148666}}}.

\bibitem[{El-Zant} \em{et~al.}(2001){El-Zant}, {Shlosman}, and
  {Hoffman}]{ElZant2001}
{El-Zant}, A.; {Shlosman}, I.; {Hoffman}, Y.
\newblock {Dark Halos: The Flattening of the Density Cusp by Dynamical
  Friction}.
\newblock {\em \apj} {\bf 2001}, {\em 560},~636--643,
  \href{http://xxx.lanl.gov/abs/astro-ph/0103386}{{\normalfont
  [astro-ph/0103386]}}.
\newblock
  doi:{\changeurlcolor{black}\href{https://doi.org/10.1086/322516}{\detokenize{10.1086/322516}}}.

\bibitem[{El-Zant} \em{et~al.}(2004){El-Zant}, {Hoffman}, {Primack}, {Combes},
  and {Shlosman}]{ElZant2004}
{El-Zant}, A.A.; {Hoffman}, Y.; {Primack}, J.; {Combes}, F.; {Shlosman}, I.
\newblock {Flat-cored Dark Matter in Cuspy Clusters of Galaxies}.
\newblock {\em \apjl} {\bf 2004}, {\em 607},~L75--L78,
  \href{http://xxx.lanl.gov/abs/astro-ph/0309412}{{\normalfont
  [astro-ph/0309412]}}.
\newblock
  doi:{\changeurlcolor{black}\href{https://doi.org/10.1086/421938}{\detokenize{10.1086/421938}}}.

\bibitem[{Ma} and {Boylan-Kolchin}(2004)]{Ma2004}
{Ma}, C.P.; {Boylan-Kolchin}, M.
\newblock {Are Halos of Collisionless Cold Dark Matter Collisionless?}
\newblock {\em Physical Review Letters} {\bf 2004}, {\em 93},~021301,
  \href{http://xxx.lanl.gov/abs/astro-ph/0403102}{{\normalfont
  [astro-ph/0403102]}}.
\newblock
  doi:{\changeurlcolor{black}\href{https://doi.org/10.1103/PhysRevLett.93.021301}{\detokenize{10.1103/PhysRevLett.93.021301}}}.

\bibitem[{Nipoti} \em{et~al.}(2004){Nipoti}, {Treu}, {Ciotti}, and
  {Stiavelli}]{Nipoti2004}
{Nipoti}, C.; {Treu}, T.; {Ciotti}, L.; {Stiavelli}, M.
\newblock {Galactic cannibalism and cold dark matter density profiles}.
\newblock {\em \mnras} {\bf 2004}, {\em 355},~1119--1124,
  \href{http://xxx.lanl.gov/abs/astro-ph/0404127}{{\normalfont
  [astro-ph/0404127]}}.
\newblock
  doi:{\changeurlcolor{black}\href{https://doi.org/10.1111/j.1365-2966.2004.08385.x}{\detokenize{10.1111/j.1365-2966.2004.08385.x}}}.

\bibitem[{Romano-D{\'{\i}}az} \em{et~al.}(2008){Romano-D{\'{\i}}az},
  {Shlosman}, {Hoffman}, and {Heller}]{RomanoDiaz2008}
{Romano-D{\'{\i}}az}, E.; {Shlosman}, I.; {Hoffman}, Y.; {Heller}, C.
\newblock {Erasing Dark Matter Cusps in Cosmological Galactic Halos with
  Baryons}.
\newblock {\em \apjl} {\bf 2008}, {\em 685},~L105--L108,
  \href{http://xxx.lanl.gov/abs/0808.0195}{{\normalfont [0808.0195]}}.
\newblock
  doi:{\changeurlcolor{black}\href{https://doi.org/10.1086/592687}{\detokenize{10.1086/592687}}}.

\bibitem[{Romano-D{\'{\i}}az} \em{et~al.}(2009){Romano-D{\'{\i}}az},
  {Shlosman}, {Heller}, and {Hoffman}]{RomanoDiaz2009}
{Romano-D{\'{\i}}az}, E.; {Shlosman}, I.; {Heller}, C.; {Hoffman}, Y.
\newblock {Dissecting Galaxy Formation. I. Comparison Between Pure Dark Matter
  and Baryonic Models}.
\newblock {\em \apj} {\bf 2009}, {\em 702},~1250--1267,
  \href{http://xxx.lanl.gov/abs/0901.1317}{{\normalfont
  [arXiv:astro-ph.CO/0901.1317]}}.
\newblock
  doi:{\changeurlcolor{black}\href{https://doi.org/10.1088/0004-637X/702/2/1250}{\detokenize{10.1088/0004-637X/702/2/1250}}}.

\bibitem[{Cole} \em{et~al.}(2011){Cole}, {Dehnen}, and {Wilkinson}]{Cole2011}
{Cole}, D.R.; {Dehnen}, W.; {Wilkinson}, M.I.
\newblock {Weakening dark matter cusps by clumpy baryonic infall}.
\newblock {\em \mnras} {\bf 2011}, {\em 416},~1118--1134,
  \href{http://xxx.lanl.gov/abs/1105.4050}{{\normalfont
  [arXiv:astro-ph.CO/1105.4050]}}.
\newblock
  doi:{\changeurlcolor{black}\href{https://doi.org/10.1111/j.1365-2966.2011.19110.x}{\detokenize{10.1111/j.1365-2966.2011.19110.x}}}.

\bibitem[{Inoue} and {Saitoh}(2011)]{Inoue2011}
{Inoue}, S.; {Saitoh}, T.R.
\newblock {Cores and revived cusps of dark matter haloes in disc galaxy
  formation through clump clusters}.
\newblock {\em \mnras} {\bf 2011}, {\em 418},~2527--2531,
  \href{http://xxx.lanl.gov/abs/1108.0906}{{\normalfont
  [arXiv:astro-ph.CO/1108.0906]}}.
\newblock
  doi:{\changeurlcolor{black}\href{https://doi.org/10.1111/j.1365-2966.2011.19873.x}{\detokenize{10.1111/j.1365-2966.2011.19873.x}}}.

\bibitem[{Nipoti} and {Binney}(2015)]{Nipoti2015}
{Nipoti}, C.; {Binney}, J.
\newblock {Early flattening of dark matter cusps in dwarf spheroidal galaxies}.
\newblock {\em \mnras} {\bf 2015}, {\em 446},~1820--1828,
  \href{http://xxx.lanl.gov/abs/1410.6169}{{\normalfont [1410.6169]}}.
\newblock
  doi:{\changeurlcolor{black}\href{https://doi.org/10.1093/mnras/stu2217}{\detokenize{10.1093/mnras/stu2217}}}.

\bibitem[{Del Popolo}(2010)]{DelPopolo2010}
{Del Popolo}, A.
\newblock {On the universality of density profiles}.
\newblock {\em \mnras} {\bf 2010}, {\em 408},~1808--1817,
  \href{http://xxx.lanl.gov/abs/1012.4322}{{\normalfont
  [arXiv:astro-ph.CO/1012.4322]}}.
\newblock
  doi:{\changeurlcolor{black}\href{https://doi.org/10.1111/j.1365-2966.2010.17288.x}{\detokenize{10.1111/j.1365-2966.2010.17288.x}}}.

\bibitem[{Di Cintio} \em{et~al.}(2014){Di Cintio}, {Brook}, {Macci{\`o}},
  {Stinson}, {Knebe}, {Dutton}, and {Wadsley}]{DiCintio2014}
{Di Cintio}, A.; {Brook}, C.B.; {Macci{\`o}}, A.V.; {Stinson}, G.S.; {Knebe},
  A.; {Dutton}, A.A.; {Wadsley}, J.
\newblock {The dependence of dark matter profiles on the stellar-to-halo mass
  ratio: a prediction for cusps versus cores}.
\newblock {\em \mnras} {\bf 2014}, {\em 437},~415--423,
  \href{http://xxx.lanl.gov/abs/1306.0898}{{\normalfont [1306.0898]}}.
\newblock
  doi:{\changeurlcolor{black}\href{https://doi.org/10.1093/mnras/stt1891}{\detokenize{10.1093/mnras/stt1891}}}.

\bibitem[{Zolotov} \em{et~al.}(2012){Zolotov}, {Brooks}, {Willman},
  {Governato}, {Pontzen}, {Christensen}, {Dekel}, {Quinn}, {Shen}, and
  {Wadsley}]{Zolotov2012}
{Zolotov}, A.; {Brooks}, A.M.; {Willman}, B.; {Governato}, F.; {Pontzen}, A.;
  {Christensen}, C.; {Dekel}, A.; {Quinn}, T.; {Shen}, S.; {Wadsley}, J.
\newblock {Baryons Matter: Why Luminous Satellite Galaxies have Reduced Central
  Masses}.
\newblock {\em \apj} {\bf 2012}, {\em 761},~71,
  \href{http://xxx.lanl.gov/abs/1207.0007}{{\normalfont
  [arXiv:astro-ph.CO/1207.0007]}}.
\newblock
  doi:{\changeurlcolor{black}\href{https://doi.org/10.1088/0004-637X/761/1/71}{\detokenize{10.1088/0004-637X/761/1/71}}}.

\bibitem[{Martizzi} \em{et~al.}(2013){Martizzi}, {Teyssier}, and
  {Moore}]{Martizzi2013}
{Martizzi}, D.; {Teyssier}, R.; {Moore}, B.
\newblock {Cusp-core transformations induced by AGN feedback in the progenitors
  of cluster galaxies}.
\newblock {\em \mnras} {\bf 2013}, {\em 432},~1947--1954,
  \href{http://xxx.lanl.gov/abs/1211.2648}{{\normalfont [1211.2648]}}.
\newblock
  doi:{\changeurlcolor{black}\href{https://doi.org/10.1093/mnras/stt297}{\detokenize{10.1093/mnras/stt297}}}.

\bibitem[{Teyssier} \em{et~al.}(2013){Teyssier}, {Pontzen}, {Dubois}, and
  {Read}]{Teyssier2013}
{Teyssier}, R.; {Pontzen}, A.; {Dubois}, Y.; {Read}, J.I.
\newblock {Cusp-core transformations in dwarf galaxies: observational
  predictions}.
\newblock {\em \mnras} {\bf 2013}, {\em 429},~3068--3078,
  \href{http://xxx.lanl.gov/abs/1206.4895}{{\normalfont
  [arXiv:astro-ph.CO/1206.4895]}}.
\newblock
  doi:{\changeurlcolor{black}\href{https://doi.org/10.1093/mnras/sts563}{\detokenize{10.1093/mnras/sts563}}}.

\bibitem[{Chan} \em{et~al.}(2015){Chan}, {Kere{\v{s}}}, {O{\~n}orbe},
  {Hopkins}, {Muratov}, {Faucher-Gigu{\`e}re}, and {Quataert}]{Chan2015}
{Chan}, T.K.; {Kere{\v{s}}}, D.; {O{\~n}orbe}, J.; {Hopkins}, P.F.; {Muratov},
  A.L.; {Faucher-Gigu{\`e}re}, C.A.; {Quataert}, E.
\newblock {The impact of baryonic physics on the structure of dark matter
  haloes: the view from the FIRE cosmological simulations}.
\newblock {\em \mnras} {\bf 2015}, {\em 454},~2981--3001,
  \href{http://xxx.lanl.gov/abs/1507.02282}{{\normalfont
  [arXiv:astro-ph.GA/1507.02282]}}.
\newblock
  doi:{\changeurlcolor{black}\href{https://doi.org/10.1093/mnras/stv2165}{\detokenize{10.1093/mnras/stv2165}}}.

\bibitem[{Tollet} \em{et~al.}(2016){Tollet}, {Macci{\`o}}, {Dutton}, {Stinson},
  {Wang}, {Penzo}, {Gutcke}, {Buck}, {Kang}, {Brook}, {Di Cintio}, {Keller},
  and {Wadsley}]{Tollet2016}
{Tollet}, E.; {Macci{\`o}}, A.V.; {Dutton}, A.A.; {Stinson}, G.S.; {Wang}, L.;
  {Penzo}, C.; {Gutcke}, T.A.; {Buck}, T.; {Kang}, X.; {Brook}, C.;  et~al.
\newblock {NIHAO - IV: core creation and destruction in dark matter density
  profiles across cosmic time}.
\newblock {\em \mnras} {\bf 2016}, {\em 456},~3542--3552,
  \href{http://xxx.lanl.gov/abs/1507.03590}{{\normalfont [1507.03590]}}.
\newblock
  doi:{\changeurlcolor{black}\href{https://doi.org/10.1093/mnras/stv2856}{\detokenize{10.1093/mnras/stv2856}}}.

\bibitem[{Peirani} \em{et~al.}(2017){Peirani}, {Dubois}, {Volonteri},
  {Devriendt}, {Bundy}, {Silk}, {Pichon}, {Kaviraj}, {Gavazzi}, and
  {Habouzit}]{Peirani2017}
{Peirani}, S.; {Dubois}, Y.; {Volonteri}, M.; {Devriendt}, J.; {Bundy}, K.;
  {Silk}, J.; {Pichon}, C.; {Kaviraj}, S.; {Gavazzi}, R.; {Habouzit}, M.
\newblock {Density profile of dark matter haloes and galaxies in the
  HORIZON-AGN simulation: the impact of AGN feedback}.
\newblock {\em \mnras} {\bf 2017}, {\em 472},~2153--2169,
  \href{http://xxx.lanl.gov/abs/1611.09922}{{\normalfont [1611.09922]}}.
\newblock
  doi:{\changeurlcolor{black}\href{https://doi.org/10.1093/mnras/stx2099}{\detokenize{10.1093/mnras/stx2099}}}.

\bibitem[{Macci{\`o}} \em{et~al.}(2020){Macci{\`o}}, {Crespi}, {Blank}, and
  {Kang}]{Maccio2020}
{Macci{\`o}}, A.V.; {Crespi}, S.; {Blank}, M.; {Kang}, X.
\newblock {NIHAO-XXIII: Dark Matter density shaped by Black Hole feedback}.
\newblock {\em \mnras} {\bf 2020},
  \href{http://xxx.lanl.gov/abs/2004.03817}{{\normalfont
  [arXiv:astro-ph.GA/2004.03817]}}.
\newblock
  doi:{\changeurlcolor{black}\href{https://doi.org/10.1093/mnrasl/slaa058}{\detokenize{10.1093/mnrasl/slaa058}}}.

\bibitem[{Dutton} \em{et~al.}(2016){Dutton}, {Macci{\`o}}, {Dekel}, {Wang},
  {Stinson}, {Obreja}, {Di Cintio}, {Brook}, {Buck}, and {Kang}]{Dutton2016}
{Dutton}, A.A.; {Macci{\`o}}, A.V.; {Dekel}, A.; {Wang}, L.; {Stinson}, G.;
  {Obreja}, A.; {Di Cintio}, A.; {Brook}, C.; {Buck}, T.; {Kang}, X.
\newblock {NIHAO IX: the role of gas inflows and outflows in driving the
  contraction and expansion of cold dark matter haloes}.
\newblock {\em \mnras} {\bf 2016}, {\em 461},~2658--2675,
  \href{http://xxx.lanl.gov/abs/1605.05323}{{\normalfont [1605.05323]}}.
\newblock
  doi:{\changeurlcolor{black}\href{https://doi.org/10.1093/mnras/stw1537}{\detokenize{10.1093/mnras/stw1537}}}.

\bibitem[{Del Popolo} \em{et~al.}(2021){Del Popolo}, {Le Delliou}, and
  {Deliyergiyev}]{DelPopolo2021}
{Del Popolo}, A.; {Le Delliou}, M.; {Deliyergiyev}, M.
\newblock {Cluster density slopes from dark matter-baryons energy transfer}.
\newblock {\em Physics of the Dark Universe} {\bf 2021}, {\em 33},~100847,
  \href{http://xxx.lanl.gov/abs/2108.07447}{{\normalfont
  [arXiv:astro-ph.GA/2108.07447]}}.
\newblock
  doi:{\changeurlcolor{black}\href{https://doi.org/10.1016/j.dark.2021.100847}{\detokenize{10.1016/j.dark.2021.100847}}}.

\bibitem[{Einasto}(1965)]{Einasto1965}
{Einasto}, J.
\newblock {On the Construction of a Composite Model for the Galaxy and on the
  Determination of the System of Galactic Parameters}.
\newblock {\em Trudy Astrofizicheskogo Instituta Alma-Ata} {\bf 1965}, {\em
  5},~87--100.

\bibitem[{Mamon} \em{et~al.}(2010){Mamon}, {Biviano}, and {Murante}]{Mamon2010}
{Mamon}, G.A.; {Biviano}, A.; {Murante}, G.
\newblock {The universal distribution of halo interlopers in projected phase
  space. Bias in galaxy cluster concentration and velocity anisotropy?}
\newblock {\em \aap} {\bf 2010}, {\em 520},~A30,
  \href{http://xxx.lanl.gov/abs/1003.0033}{{\normalfont
  [arXiv:astro-ph.CO/1003.0033]}}.
\newblock
  doi:{\changeurlcolor{black}\href{https://doi.org/10.1051/0004-6361/200913948}{\detokenize{10.1051/0004-6361/200913948}}}.

\bibitem[{Retana-Montenegro} \em{et~al.}(2012){Retana-Montenegro}, {van Hese},
  {Gentile}, {Baes}, and {Frutos-Alfaro}]{Retana-Montenegro2012}
{Retana-Montenegro}, E.; {van Hese}, E.; {Gentile}, G.; {Baes}, M.;
  {Frutos-Alfaro}, F.
\newblock {Analytical properties of Einasto dark matter haloes}.
\newblock {\em \aap} {\bf 2012}, {\em 540},~A70,
  \href{http://xxx.lanl.gov/abs/1202.5242}{{\normalfont
  [arXiv:astro-ph.CO/1202.5242]}}.
\newblock
  doi:{\changeurlcolor{black}\href{https://doi.org/10.1051/0004-6361/201118543}{\detokenize{10.1051/0004-6361/201118543}}}.

\bibitem[{An} and {Zhao}(2013)]{An2013}
{An}, J.; {Zhao}, H.
\newblock {Fitting functions for dark matter density profiles}.
\newblock {\em \mnras} {\bf 2013}, {\em 428},~2805--2811,
  \href{http://xxx.lanl.gov/abs/1209.6220}{{\normalfont
  [arXiv:astro-ph.CO/1209.6220]}}.
\newblock
  doi:{\changeurlcolor{black}\href{https://doi.org/10.1093/mnras/sts175}{\detokenize{10.1093/mnras/sts175}}}.

\bibitem[{Dekel} \em{et~al.}(2017){Dekel}, {Ishai}, {Dutton}, and
  {Maccio}]{Dekel2017}
{Dekel}, A.; {Ishai}, G.; {Dutton}, A.A.; {Maccio}, A.V.
\newblock {Dark-matter halo profiles of a general cusp/core with analytic
  velocity and potential}.
\newblock {\em \mnras} {\bf 2017}, {\em 468},~1005--1022,
  \href{http://xxx.lanl.gov/abs/1610.00916}{{\normalfont [1610.00916]}}.
\newblock
  doi:{\changeurlcolor{black}\href{https://doi.org/10.1093/mnras/stx486}{\detokenize{10.1093/mnras/stx486}}}.

\bibitem[{Zhao}(1996)]{Zhao1996}
{Zhao}, H.
\newblock {Analytical models for galactic nuclei}.
\newblock {\em \mnras} {\bf 1996}, {\em 278},~488--496,
  \href{http://xxx.lanl.gov/abs/astro-ph/9509122}{{\normalfont
  [astro-ph/9509122]}}.

\bibitem[{Freundlich} \em{et~al.}(2020){Freundlich}, {Jiang}, {Dekel},
  {Cornuault}, {Ginzburg}, {Koskas}, {Lapiner}, {Dutton}, and
  {Macci{\`o}}]{Freundlich2020}
{Freundlich}, J.; {Jiang}, F.; {Dekel}, A.; {Cornuault}, N.; {Ginzburg}, O.;
  {Koskas}, R.; {Lapiner}, S.; {Dutton}, A.; {Macci{\`o}}, A.V.
\newblock {The Dekel-Zhao profile: a mass-dependent dark-matter density profile
  with flexible inner slope and analytic potential, velocity dispersion, and
  lensing properties}.
\newblock {\em \mnras} {\bf 2020}, {\em 499},~2912--2933,
  \href{http://xxx.lanl.gov/abs/2004.08395}{{\normalfont
  [arXiv:astro-ph.GA/2004.08395]}}.
\newblock
  doi:{\changeurlcolor{black}\href{https://doi.org/10.1093/mnras/staa2790}{\detokenize{10.1093/mnras/staa2790}}}.

\bibitem[{Lazar} \em{et~al.}(2020){Lazar}, {Bullock}, {Boylan-Kolchin}, {Chan},
  {Hopkins}, {Graus}, {Wetzel}, {El-Badry}, {Wheeler}, {Straight},
  {Kere{\v{s}}}, {Faucher-Gigu{\`e}re}, {Fitts}, and
  {Garrison-Kimmel}]{Lazar2020}
{Lazar}, A.; {Bullock}, J.S.; {Boylan-Kolchin}, M.; {Chan}, T.K.; {Hopkins},
  P.F.; {Graus}, A.S.; {Wetzel}, A.; {El-Badry}, K.; {Wheeler}, C.; {Straight},
  M.C.;  et~al.
\newblock {A dark matter profile to model diverse feedback-induced core sizes
  of {\ensuremath{\Lambda}}CDM haloes}.
\newblock {\em \mnras} {\bf 2020}, {\em 497},~2393--2417,
  \href{http://xxx.lanl.gov/abs/2004.10817}{{\normalfont
  [arXiv:astro-ph.GA/2004.10817]}}.
\newblock
  doi:{\changeurlcolor{black}\href{https://doi.org/10.1093/mnras/staa2101}{\detokenize{10.1093/mnras/staa2101}}}.

\bibitem[{Read} \em{et~al.}(2016){Read}, {Agertz}, and {Collins}]{Read2016}
{Read}, J.I.; {Agertz}, O.; {Collins}, M.L.M.
\newblock {Dark matter cores all the way down}.
\newblock {\em \mnras} {\bf 2016}, {\em 459},~2573--2590,
  \href{http://xxx.lanl.gov/abs/1508.04143}{{\normalfont
  [arXiv:astro-ph.GA/1508.04143]}}.
\newblock
  doi:{\changeurlcolor{black}\href{https://doi.org/10.1093/mnras/stw713}{\detokenize{10.1093/mnras/stw713}}}.

\bibitem[{Di Cintio} \em{et~al.}(2014){Di Cintio}, {Brook}, {Dutton},
  {Macci{\`o}}, {Stinson}, and {Knebe}]{DiCintio2014b}
{Di Cintio}, A.; {Brook}, C.B.; {Dutton}, A.A.; {Macci{\`o}}, A.V.; {Stinson},
  G.S.; {Knebe}, A.
\newblock {A mass-dependent density profile for dark matter haloes including
  the influence of galaxy formation}.
\newblock {\em \mnras} {\bf 2014}, {\em 441},~2986--2995,
  \href{http://xxx.lanl.gov/abs/1404.5959}{{\normalfont [1404.5959]}}.
\newblock
  doi:{\changeurlcolor{black}\href{https://doi.org/10.1093/mnras/stu729}{\detokenize{10.1093/mnras/stu729}}}.

\bibitem[{Del Popolo} and {Kroupa}(2009)]{DelPopolo2009a}
{Del Popolo}, A.; {Kroupa}, P.
\newblock {Density profiles of dark matter haloes on galactic and cluster
  scales}.
\newblock {\em \aap} {\bf 2009}, {\em 502},~733--747,
  \href{http://xxx.lanl.gov/abs/0906.1146}{{\normalfont
  [arXiv:astro-ph.CO/0906.1146]}}.
\newblock
  doi:{\changeurlcolor{black}\href{https://doi.org/10.1051/0004-6361/200811404}{\detokenize{10.1051/0004-6361/200811404}}}.

\bibitem[{Gunn} and {Gott}(1972)]{Gunn1972}
{Gunn}, J.E.; {Gott}, III, J.R.
\newblock {On the Infall of Matter Into Clusters of Galaxies and Some Effects
  on Their Evolution}.
\newblock {\em \apj} {\bf 1972}, {\em 176},~1.
\newblock
  doi:{\changeurlcolor{black}\href{https://doi.org/10.1086/151605}{\detokenize{10.1086/151605}}}.

\bibitem[{Bertschinger}(1985)]{Bertschinger1985}
{Bertschinger}, E.
\newblock {Self-similar secondary infall and accretion in an Einstein-de Sitter
  universe}.
\newblock {\em \apjs} {\bf 1985}, {\em 58},~39--65.
\newblock
  doi:{\changeurlcolor{black}\href{https://doi.org/10.1086/191028}{\detokenize{10.1086/191028}}}.

\bibitem[{Hoffman} and {Shaham}(1985)]{Hoffman1985}
{Hoffman}, Y.; {Shaham}, J.
\newblock {Local density maxima - Progenitors of structure}.
\newblock {\em \apj} {\bf 1985}, {\em 297},~16--22.
\newblock
  doi:{\changeurlcolor{black}\href{https://doi.org/10.1086/163498}{\detokenize{10.1086/163498}}}.

\bibitem[{Ryden} and {Gunn}(1987)]{Ryden1987}
{Ryden}, B.S.; {Gunn}, J.E.
\newblock {Galaxy formation by gravitational collapse}.
\newblock {\em \apj} {\bf 1987}, {\em 318},~15--31.
\newblock
  doi:{\changeurlcolor{black}\href{https://doi.org/10.1086/165349}{\detokenize{10.1086/165349}}}.

\bibitem[{Ascasibar} \em{et~al.}(2004){Ascasibar}, {Yepes}, {Gottl{\"o}ber},
  and {M{\"u}ller}]{Ascasibar2004}
{Ascasibar}, Y.; {Yepes}, G.; {Gottl{\"o}ber}, S.; {M{\"u}ller}, V.
\newblock {On the physical origin of dark matter density profiles}.
\newblock {\em \mnras} {\bf 2004}, {\em 352},~1109--1120,
  \href{http://xxx.lanl.gov/abs/arXiv:astro-ph/0312221}{{\normalfont
  [arXiv:astro-ph/0312221]}}.
\newblock
  doi:{\changeurlcolor{black}\href{https://doi.org/10.1111/j.1365-2966.2004.08005.x}{\detokenize{10.1111/j.1365-2966.2004.08005.x}}}.

\bibitem[{Williams} \em{et~al.}(2004){Williams}, {Babul}, and
  {Dalcanton}]{Williams2004}
{Williams}, L.L.R.; {Babul}, A.; {Dalcanton}, J.J.
\newblock {Investigating the Origins of Dark Matter Halo Density Profiles}.
\newblock {\em \apj} {\bf 2004}, {\em 604},~18--39,
  \href{http://xxx.lanl.gov/abs/astro-ph/0312002}{{\normalfont
  [astro-ph/0312002]}}.
\newblock
  doi:{\changeurlcolor{black}\href{https://doi.org/10.1086/381722}{\detokenize{10.1086/381722}}}.

\bibitem[{Ryden}(1988)]{Ryden1988}
{Ryden}, B.S.
\newblock {Galaxy formation - The role of tidal torques and dissipational
  infall}.
\newblock {\em \apj} {\bf 1988}, {\em 329},~589--611.
\newblock
  doi:{\changeurlcolor{black}\href{https://doi.org/10.1086/166406}{\detokenize{10.1086/166406}}}.

\bibitem[{Del Popolo} and {Gambera}(1997)]{DelPopolo1997}
{Del Popolo}, A.; {Gambera}, M.
\newblock {Substructure effects on the collapse of density perturbations.}
\newblock {\em \aap} {\bf 1997}, {\em 321},~691--695,
  \href{http://xxx.lanl.gov/abs/astro-ph/9610052}{{\normalfont
  [astro-ph/9610052]}}.

\bibitem[{Del Popolo} and {Gambera}(2000)]{DelPopolo2000}
{Del Popolo}, A.; {Gambera}, M.
\newblock {Non radial motions and the shapes and the abundance of clusters of
  galaxies}.
\newblock {\em \aap} {\bf 2000}, {\em 357},~809--815,
  \href{http://xxx.lanl.gov/abs/astro-ph/9909156}{{\normalfont
  [astro-ph/9909156]}}.

\bibitem[{Blumenthal} \em{et~al.}(1986){Blumenthal}, {Faber}, {Flores}, and
  {Primack}]{Blumenthal1986}
{Blumenthal}, G.R.; {Faber}, S.M.; {Flores}, R.; {Primack}, J.R.
\newblock {Contraction of dark matter galactic halos due to baryonic infall}.
\newblock {\em \apj} {\bf 1986}, {\em 301},~27--34.
\newblock
  doi:{\changeurlcolor{black}\href{https://doi.org/10.1086/163867}{\detokenize{10.1086/163867}}}.

\bibitem[{Gnedin} \em{et~al.}(2004){Gnedin}, {Kravtsov}, {Klypin}, and
  {Nagai}]{Gnedin2004}
{Gnedin}, O.Y.; {Kravtsov}, A.V.; {Klypin}, A.A.; {Nagai}, D.
\newblock {Response of Dark Matter Halos to Condensation of Baryons:
  Cosmological Simulations and Improved Adiabatic Contraction Model}.
\newblock {\em \apj} {\bf 2004}, {\em 616},~16--26,
  \href{http://xxx.lanl.gov/abs/astro-ph/0406247}{{\normalfont
  [astro-ph/0406247]}}.
\newblock
  doi:{\changeurlcolor{black}\href{https://doi.org/10.1086/424914}{\detokenize{10.1086/424914}}}.

\bibitem[{Klypin} \em{et~al.}(2002){Klypin}, {Zhao}, and
  {Somerville}]{Klypin2002}
{Klypin}, A.; {Zhao}, H.; {Somerville}, R.S.
\newblock {{$\Lambda$}CDM-based Models for the Milky Way and M31. I. Dynamical
  Models}.
\newblock {\em \apj} {\bf 2002}, {\em 573},~597--613,
  \href{http://xxx.lanl.gov/abs/astro-ph/0110390}{{\normalfont
  [astro-ph/0110390]}}.
\newblock
  doi:{\changeurlcolor{black}\href{https://doi.org/10.1086/340656}{\detokenize{10.1086/340656}}}.

\bibitem[{Gustafsson} \em{et~al.}(2006){Gustafsson}, {Fairbairn}, and
  {Sommer-Larsen}]{Gustafsson2006}
{Gustafsson}, M.; {Fairbairn}, M.; {Sommer-Larsen}, J.
\newblock {Baryonic pinching of galactic dark matter halos}.
\newblock {\em \prd} {\bf 2006}, {\em 74},~123522,
  \href{http://xxx.lanl.gov/abs/astro-ph/0608634}{{\normalfont
  [astro-ph/0608634]}}.
\newblock
  doi:{\changeurlcolor{black}\href{https://doi.org/10.1103/PhysRevD.74.123522}{\detokenize{10.1103/PhysRevD.74.123522}}}.

\bibitem[{De Lucia} and {Helmi}(2008)]{DeLucia2008}
{De Lucia}, G.; {Helmi}, A.
\newblock {The Galaxy and its stellar halo: insights on their formation from a
  hybrid cosmological approach}.
\newblock {\em \mnras} {\bf 2008}, {\em 391},~14--31,
  \href{http://xxx.lanl.gov/abs/0804.2465}{{\normalfont [0804.2465]}}.
\newblock
  doi:{\changeurlcolor{black}\href{https://doi.org/10.1111/j.1365-2966.2008.13862.x}{\detokenize{10.1111/j.1365-2966.2008.13862.x}}}.

\bibitem[{Li} \em{et~al.}(2010){Li}, {De Lucia}, and {Helmi}]{Li2010}
{Li}, Y.S.; {De Lucia}, G.; {Helmi}, A.
\newblock {On the nature of the Milky Way satellites}.
\newblock {\em \mnras} {\bf 2010}, {\em 401},~2036--2052,
  \href{http://xxx.lanl.gov/abs/0909.1291}{{\normalfont
  [arXiv:astro-ph.GA/0909.1291]}}.
\newblock
  doi:{\changeurlcolor{black}\href{https://doi.org/10.1111/j.1365-2966.2009.15803.x}{\detokenize{10.1111/j.1365-2966.2009.15803.x}}}.

\bibitem[{Martizzi} \em{et~al.}(2012){Martizzi}, {Teyssier}, {Moore}, and
  {Wentz}]{Martizzi2012}
{Martizzi}, D.; {Teyssier}, R.; {Moore}, B.; {Wentz}, T.
\newblock {The effects of baryon physics, black holes and active galactic
  nucleus feedback on the mass distribution in clusters of galaxies}.
\newblock {\em \mnras} {\bf 2012}, {\em 422},~3081--3091,
  \href{http://xxx.lanl.gov/abs/1112.2752}{{\normalfont
  [arXiv:astro-ph.CO/1112.2752]}}.
\newblock
  doi:{\changeurlcolor{black}\href{https://doi.org/10.1111/j.1365-2966.2012.20879.x}{\detokenize{10.1111/j.1365-2966.2012.20879.x}}}.

\bibitem[{Del Popolo} \em{et~al.}(2013{\natexlab{a}}){Del Popolo}, {Pace}, and
  {Lima}]{DelPopolo2013a}
{Del Popolo}, A.; {Pace}, F.; {Lima}, J.A.S.
\newblock {Extended Spherical Collapse and the Accelerating Universe}.
\newblock {\em International Journal of Modern Physics D} {\bf 2013}, {\em
  22},~50038,  \href{http://xxx.lanl.gov/abs/1207.5789}{{\normalfont
  [arXiv:astro-ph.CO/1207.5789]}}.
\newblock
  doi:{\changeurlcolor{black}\href{https://doi.org/10.1142/S0218271813500387}{\detokenize{10.1142/S0218271813500387}}}.

\bibitem[{Del Popolo} \em{et~al.}(2013{\natexlab{b}}){Del Popolo}, {Pace}, and
  {Lima}]{DelPopolo2013b}
{Del Popolo}, A.; {Pace}, F.; {Lima}, J.A.S.
\newblock {Spherical collapse model with shear and angular momentum in dark
  energy cosmologies}.
\newblock {\em \mnras} {\bf 2013}, {\em 430},~628--637,
  \href{http://xxx.lanl.gov/abs/1212.5092}{{\normalfont
  [arXiv:astro-ph.CO/1212.5092]}}.
\newblock
  doi:{\changeurlcolor{black}\href{https://doi.org/10.1093/mnras/sts669}{\detokenize{10.1093/mnras/sts669}}}.

\bibitem[{Del Popolo} \em{et~al.}(2013{\natexlab{c}}){Del Popolo}, {Pace},
  {Maydanyuk}, {Lima}, and {Jesus}]{DelPopolo2013c}
{Del Popolo}, A.; {Pace}, F.; {Maydanyuk}, S.P.; {Lima}, J.A.S.; {Jesus}, J.F.
\newblock {Shear and rotation in Chaplygin cosmology}.
\newblock {\em \prd} {\bf 2013}, {\em 87},~043527,
  \href{http://xxx.lanl.gov/abs/1303.3628}{{\normalfont
  [arXiv:astro-ph.CO/1303.3628]}}.
\newblock
  doi:{\changeurlcolor{black}\href{https://doi.org/10.1103/PhysRevD.87.043527}{\detokenize{10.1103/PhysRevD.87.043527}}}.

\bibitem[{Del Popolo} and {Pace}(2016)]{DelPopolo2016a}
{Del Popolo}, A.; {Pace}, F.
\newblock {The Cusp/Core problem: supernovae feedback versus the baryonic
  clumps and dynamical friction model}.
\newblock {\em \apss} {\bf 2016}, {\em 361},~162,
  \href{http://xxx.lanl.gov/abs/1502.01947}{{\normalfont [1502.01947]}}.
\newblock
  doi:{\changeurlcolor{black}\href{https://doi.org/10.1007/s10509-016-2742-z}{\detokenize{10.1007/s10509-016-2742-z}}}.

\bibitem[{Del Popolo}(2016)]{DelPopolo2016b}
{Del Popolo}, A.
\newblock {On the dark matter haloes inner structure and galaxy morphology}.
\newblock {\em \apss} {\bf 2016}, {\em 361},~222,
  \href{http://xxx.lanl.gov/abs/1607.07408}{{\normalfont [1607.07408]}}.
\newblock
  doi:{\changeurlcolor{black}\href{https://doi.org/10.1007/s10509-016-2810-4}{\detokenize{10.1007/s10509-016-2810-4}}}.

\bibitem[{Del Popolo}(2011)]{DelPopolo2011}
{Del Popolo}, A.
\newblock {Non-power law behavior of the radial profile of phase-space density
  of halos}.
\newblock {\em \jcap} {\bf 2011}, {\em 7},~14,
  \href{http://xxx.lanl.gov/abs/1112.4185}{{\normalfont
  [arXiv:astro-ph.CO/1112.4185]}}.
\newblock
  doi:{\changeurlcolor{black}\href{https://doi.org/10.1088/1475-7516/2011/07/014}{\detokenize{10.1088/1475-7516/2011/07/014}}}.

\bibitem[{Del Popolo} \em{et~al.}(2013){Del Popolo}, {Cardone}, and
  {Belvedere}]{DelPopolo2013d}
{Del Popolo}, A.; {Cardone}, V.F.; {Belvedere}, G.
\newblock {Surface density of dark matter haloes on galactic and cluster
  scales}.
\newblock {\em \mnras} {\bf 2013}, {\em 429},~1080--1087,
  \href{http://xxx.lanl.gov/abs/1212.6797}{{\normalfont
  [arXiv:astro-ph.CO/1212.6797]}}.
\newblock
  doi:{\changeurlcolor{black}\href{https://doi.org/10.1093/mnras/sts389}{\detokenize{10.1093/mnras/sts389}}}.

\bibitem[{Pontzen} and {Governato}(2012)]{Pontzen2012}
{Pontzen}, A.; {Governato}, F.
\newblock {How supernova feedback turns dark matter cusps into cores}.
\newblock {\em \mnras} {\bf 2012}, {\em 421},~3464--3471,
  \href{http://xxx.lanl.gov/abs/1106.0499}{{\normalfont
  [arXiv:astro-ph.CO/1106.0499]}}.
\newblock
  doi:{\changeurlcolor{black}\href{https://doi.org/10.1111/j.1365-2966.2012.20571.x}{\detokenize{10.1111/j.1365-2966.2012.20571.x}}}.

\bibitem[{Gunn}(1977)]{Gunn1977}
{Gunn}, J.E.
\newblock {Massive galactic halos. I - Formation and evolution}.
\newblock {\em \apj} {\bf 1977}, {\em 218},~592--598.
\newblock
  doi:{\changeurlcolor{black}\href{https://doi.org/10.1086/155715}{\detokenize{10.1086/155715}}}.

\bibitem[{Fillmore} and {Goldreich}(1984)]{Fillmore1984}
{Fillmore}, J.A.; {Goldreich}, P.
\newblock {Self-similar gravitational collapse in an expanding universe}.
\newblock {\em \apj} {\bf 1984}, {\em 281},~1--8.
\newblock
  doi:{\changeurlcolor{black}\href{https://doi.org/10.1086/162070}{\detokenize{10.1086/162070}}}.

\bibitem[{Komatsu} \em{et~al.}(2009){Komatsu}, {Dunkley}, {Nolta}, and {et
  al.}]{Komatsu2009}
{Komatsu}, E.; {Dunkley}, J.; {Nolta}, M.R.; {et al.}.
\newblock {Five-Year Wilkinson Microwave Anisotropy Probe Observations:
  Cosmological Interpretation}.
\newblock {\em \apjs} {\bf 2009}, {\em 180},~330--376,
  \href{http://xxx.lanl.gov/abs/0803.0547}{{\normalfont [0803.0547]}}.
\newblock
  doi:{\changeurlcolor{black}\href{https://doi.org/10.1088/0067-0049/180/2/330}{\detokenize{10.1088/0067-0049/180/2/330}}}.

\bibitem[{Komatsu} \em{et~al.}(2011){Komatsu}, {Smith}, {Dunkley}, and
  {et~al.}]{Komatsu2011}
{Komatsu}, E.; {Smith}, K.M.; {Dunkley}, J.; {et~al.}.
\newblock {Seven-year Wilkinson Microwave Anisotropy Probe (WMAP) Observations:
  Cosmological Interpretation}.
\newblock {\em \apjs} {\bf 2011}, {\em 192},~18--+,
  \href{http://xxx.lanl.gov/abs/1001.4538}{{\normalfont
  [arXiv:astro-ph.CO/1001.4538]}}.
\newblock
  doi:{\changeurlcolor{black}\href{https://doi.org/10.1088/0067-0049/192/2/18}{\detokenize{10.1088/0067-0049/192/2/18}}}.

\bibitem[{Hoyle}(1953)]{Hoyle1953}
{Hoyle}, F.
\newblock {On the Fragmentation of Gas Clouds Into Galaxies and Stars.}
\newblock {\em \apj} {\bf 1953}, {\em 118},~513.
\newblock
  doi:{\changeurlcolor{black}\href{https://doi.org/10.1086/145780}{\detokenize{10.1086/145780}}}.

\bibitem[{Peebles}(1969)]{Peebles1969}
{Peebles}, P.J.E.
\newblock {Origin of the Angular Momentum of Galaxies}.
\newblock {\em \apj} {\bf 1969}, {\em 155},~393.
\newblock
  doi:{\changeurlcolor{black}\href{https://doi.org/10.1086/149876}{\detokenize{10.1086/149876}}}.

\bibitem[{White}(1984)]{White1984}
{White}, S.D.M.
\newblock {Angular momentum growth in protogalaxies}.
\newblock {\em \apj} {\bf 1984}, {\em 286},~38--41.
\newblock
  doi:{\changeurlcolor{black}\href{https://doi.org/10.1086/162573}{\detokenize{10.1086/162573}}}.

\bibitem[{Eisenstein} and {Loeb}(1995)]{Eisenstein1995}
{Eisenstein}, D.J.; {Loeb}, A.
\newblock {An analytical model for the triaxial collapse of cosmological
  perturbations}.
\newblock {\em \apj} {\bf 1995}, {\em 439},~520--541,
  \href{http://xxx.lanl.gov/abs/arXiv:astro-ph/9405012}{{\normalfont
  [arXiv:astro-ph/9405012]}}.
\newblock
  doi:{\changeurlcolor{black}\href{https://doi.org/10.1086/175193}{\detokenize{10.1086/175193}}}.

\bibitem[{Avila-Reese} \em{et~al.}(1998){Avila-Reese}, {Firmani}, and
  {Hern{\'a}ndez}]{AvilaReese1998}
{Avila-Reese}, V.; {Firmani}, C.; {Hern{\'a}ndez}, X.
\newblock {On the Formation and Evolution of Disk Galaxies: Cosmological
  Initial Conditions and the Gravitational Collapse}.
\newblock {\em \apj} {\bf 1998}, {\em 505},~37--49,
  \href{http://xxx.lanl.gov/abs/astro-ph/9710201}{{\normalfont
  [astro-ph/9710201]}}.
\newblock
  doi:{\changeurlcolor{black}\href{https://doi.org/10.1086/306136}{\detokenize{10.1086/306136}}}.

\bibitem[{Toomre}(1964)]{Toomre1964}
{Toomre}, A.
\newblock {On the gravitational stability of a disk of stars}.
\newblock {\em \apj} {\bf 1964}, {\em 139},~1217--1238.
\newblock
  doi:{\changeurlcolor{black}\href{https://doi.org/10.1086/147861}{\detokenize{10.1086/147861}}}.

\bibitem[{Binney} and {Tremaine}(1987)]{BinneyTremaine1987}
{Binney}, J.; {Tremaine}, S.
\newblock {\em {Galactic dynamics}}; Princeton University Press: Princeton, NJ,
   1987.

\bibitem[{Krumholz} and {Dekel}(2010)]{Krumholz2010}
{Krumholz}, M.R.; {Dekel}, A.
\newblock {Survival of star-forming giant clumps in high-redshift galaxies}.
\newblock {\em \mnras} {\bf 2010}, {\em 406},~112--120,
  \href{http://xxx.lanl.gov/abs/1001.0765}{{\normalfont [1001.0765]}}.
\newblock
  doi:{\changeurlcolor{black}\href{https://doi.org/10.1111/j.1365-2966.2010.16675.x}{\detokenize{10.1111/j.1365-2966.2010.16675.x}}}.

\bibitem[{Dekel} \em{et~al.}(2009){Dekel}, {Sari}, and {Ceverino}]{Dekel2009}
{Dekel}, A.; {Sari}, R.; {Ceverino}, D.
\newblock {Formation of Massive Galaxies at High Redshift: Cold Streams, Clumpy
  Disks, and Compact Spheroids}.
\newblock {\em \apj} {\bf 2009}, {\em 703},~785--801,
  \href{http://xxx.lanl.gov/abs/0901.2458}{{\normalfont
  [arXiv:astro-ph.GA/0901.2458]}}.
\newblock
  doi:{\changeurlcolor{black}\href{https://doi.org/10.1088/0004-637X/703/1/785}{\detokenize{10.1088/0004-637X/703/1/785}}}.

\bibitem[{Ceverino} \em{et~al.}(2012){Ceverino}, {Dekel}, {Mandelker},
  {Bournaud}, {Burkert}, {Genzel}, and {Primack}]{Ceverino2012}
{Ceverino}, D.; {Dekel}, A.; {Mandelker}, N.; {Bournaud}, F.; {Burkert}, A.;
  {Genzel}, R.; {Primack}, J.
\newblock {Rotational support of giant clumps in high-z disc galaxies}.
\newblock {\em \mnras} {\bf 2012}, {\em 420},~3490--3520,
  \href{http://xxx.lanl.gov/abs/1106.5587}{{\normalfont [1106.5587]}}.
\newblock
  doi:{\changeurlcolor{black}\href{https://doi.org/10.1111/j.1365-2966.2011.20296.x}{\detokenize{10.1111/j.1365-2966.2011.20296.x}}}.

\bibitem[{Del Popolo} and {Le Delliou}(2014)]{DelPopolo2014d}
{Del Popolo}, A.; {Le Delliou}, M.
\newblock {A unified solution to the small scale problems of the {$\Lambda$}CDM
  model II: introducing parent-satellite interaction}.
\newblock {\em \jcap} {\bf 2014}, {\em 12},~51,
  \href{http://xxx.lanl.gov/abs/1408.4893}{{\normalfont [1408.4893]}}.
\newblock
  doi:{\changeurlcolor{black}\href{https://doi.org/10.1088/1475-7516/2014/12/051}{\detokenize{10.1088/1475-7516/2014/12/051}}}.

\bibitem[{Ceverino} \em{et~al.}(2010){Ceverino}, {Dekel}, and
  {Bournaud}]{Ceverino2010}
{Ceverino}, D.; {Dekel}, A.; {Bournaud}, F.
\newblock {High-redshift clumpy discs and bulges in cosmological simulations}.
\newblock {\em \mnras} {\bf 2010}, {\em 404},~2151--2169,
  \href{http://xxx.lanl.gov/abs/0907.3271}{{\normalfont
  [arXiv:astro-ph.CO/0907.3271]}}.
\newblock
  doi:{\changeurlcolor{black}\href{https://doi.org/10.1111/j.1365-2966.2010.16433.x}{\detokenize{10.1111/j.1365-2966.2010.16433.x}}}.

\bibitem[{Perez} \em{et~al.}(2013){Perez}, {Valenzuela}, {Tissera}, and
  {Michel-Dansac}]{Perez2013}
{Perez}, J.; {Valenzuela}, O.; {Tissera}, P.B.; {Michel-Dansac}, L.
\newblock {Clumpy disc and bulge formation}.
\newblock {\em \mnras} {\bf 2013}, {\em 436},~259--265,
  \href{http://xxx.lanl.gov/abs/1308.4396}{{\normalfont [1308.4396]}}.
\newblock
  doi:{\changeurlcolor{black}\href{https://doi.org/10.1093/mnras/stt1563}{\detokenize{10.1093/mnras/stt1563}}}.

\bibitem[Perret \em{et~al.}(2014)Perret, Renaud, Epinat, Amram, Bournaud,
  Contini, Teyssier, and Lambert]{Perret2013}
Perret, V.; Renaud, F.; Epinat, B.; Amram, P.; Bournaud, F.; Contini, T.;
  Teyssier, R.; Lambert, J.C.
\newblock {Evolution of the mass, size, and star formation rate in high
  redshift merging galaxies - MIRAGE --- A new sample of simulations with
  detailed stellar feedback}.
\newblock {\em Astron. Astrophys.} {\bf 2014}, {\em 562},~A1,
  \href{http://xxx.lanl.gov/abs/1307.7130}{{\normalfont
  [arXiv:astro-ph.CO/1307.7130]}}.
\newblock
  doi:{\changeurlcolor{black}\href{https://doi.org/10.1051/0004-6361/201322395}{\detokenize{10.1051/0004-6361/201322395}}}.

\bibitem[Ceverino \em{et~al.}(2014)Ceverino, Klypin, Klimek, Trujillo-Gomez,
  Churchill, and Primack]{Ceverino2013}
Ceverino, D.; Klypin, A.; Klimek, E.; Trujillo-Gomez, S.; Churchill, C.W.;
  Primack, J.
\newblock {Radiative feedback and the low efficiency of galaxy formation in
  low-mass haloes at high redshift}.
\newblock {\em Mon. Not. Roy. Astron. Soc.} {\bf 2014}, {\em 442},~1545--1559,
  \href{http://xxx.lanl.gov/abs/1307.0943}{{\normalfont
  [arXiv:astro-ph.CO/1307.0943]}}.
\newblock
  doi:{\changeurlcolor{black}\href{https://doi.org/10.1093/mnras/stu956}{\detokenize{10.1093/mnras/stu956}}}.

\bibitem[Ceverino \em{et~al.}(2015)Ceverino, Dekel, Tweed, and
  Primack]{Ceverino2014}
Ceverino, D.; Dekel, A.; Tweed, D.; Primack, J.
\newblock {Early formation of massive, compact, spheroidal galaxies with
  classical profiles by violent disc instability or mergers}.
\newblock {\em Mon. Not. Roy. Astron. Soc.} {\bf 2015}, {\em 447},~3291,
  \href{http://xxx.lanl.gov/abs/1409.2622}{{\normalfont
  [arXiv:astro-ph.GA/1409.2622]}}.
\newblock
  doi:{\changeurlcolor{black}\href{https://doi.org/10.1093/mnras/stu2694}{\detokenize{10.1093/mnras/stu2694}}}.

\bibitem[{Bournaud} \em{et~al.}(2014){Bournaud}, {Perret}, {Renaud}, {Dekel},
  {Elmegreen}, {Elmegreen}, {Teyssier}, {Amram}, {Daddi}, {Duc}, {Elbaz},
  {Epinat}, {Gabor}, {Juneau}, {Kraljic}, and {Le Floch'}]{Bournaud2014}
{Bournaud}, F.; {Perret}, V.; {Renaud}, F.; {Dekel}, A.; {Elmegreen}, B.G.;
  {Elmegreen}, D.M.; {Teyssier}, R.; {Amram}, P.; {Daddi}, E.; {Duc}, P.A.;
  et~al.
\newblock {The Long Lives of Giant Clumps and the Birth of Outflows in Gas-rich
  Galaxies at High Redshift}.
\newblock {\em \apj} {\bf 2014}, {\em 780},~57,
  \href{http://xxx.lanl.gov/abs/1307.7136}{{\normalfont
  [arXiv:astro-ph.CO/1307.7136]}}.
\newblock
  doi:{\changeurlcolor{black}\href{https://doi.org/10.1088/0004-637X/780/1/57}{\detokenize{10.1088/0004-637X/780/1/57}}}.

\bibitem[{Behrendt} \em{et~al.}(2016){Behrendt}, {Burkert}, and
  {Schartmann}]{Behrendt2016}
{Behrendt}, M.; {Burkert}, A.; {Schartmann}, M.
\newblock {Clusters of Small Clumps Can Explain the Peculiar Properties of
  Giant Clumps in High-redshift Galaxies}.
\newblock {\em \apjl} {\bf 2016}, {\em 819},~L2,
  \href{http://xxx.lanl.gov/abs/1512.03430}{{\normalfont [1512.03430]}}.
\newblock
  doi:{\changeurlcolor{black}\href{https://doi.org/10.3847/2041-8205/819/1/L2}{\detokenize{10.3847/2041-8205/819/1/L2}}}.

\bibitem[{Elmegreen} \em{et~al.}(2004){Elmegreen}, {Elmegreen}, and
  {Hirst}]{Elmegreen2004}
{Elmegreen}, D.M.; {Elmegreen}, B.G.; {Hirst}, A.C.
\newblock {Discovery of Face-on Counterparts of Chain Galaxies in the Tadpole
  Advanced Camera for Surveys Field}.
\newblock {\em \apjl} {\bf 2004}, {\em 604},~L21--L23,
  \href{http://xxx.lanl.gov/abs/astro-ph/0402477}{{\normalfont
  [astro-ph/0402477]}}.
\newblock
  doi:{\changeurlcolor{black}\href{https://doi.org/10.1086/383312}{\detokenize{10.1086/383312}}}.

\bibitem[{Elmegreen} \em{et~al.}(2009){Elmegreen}, {Elmegreen}, {Marcus},
  {Shahinyan}, {Yau}, and {Petersen}]{Elmegreen2009}
{Elmegreen}, D.M.; {Elmegreen}, B.G.; {Marcus}, M.T.; {Shahinyan}, K.; {Yau},
  A.; {Petersen}, M.
\newblock {Clumpy Galaxies in Goods and Gems: Massive Analogs of Local Dwarf
  Irregulars}.
\newblock {\em \apj} {\bf 2009}, {\em 701},~306--329,
  \href{http://xxx.lanl.gov/abs/0906.2660}{{\normalfont
  [arXiv:astro-ph.CO/0906.2660]}}.
\newblock
  doi:{\changeurlcolor{black}\href{https://doi.org/10.1088/0004-637X/701/1/306}{\detokenize{10.1088/0004-637X/701/1/306}}}.

\bibitem[{Genzel} \em{et~al.}(2011){Genzel}, {Newman}, {Jones}, {F{\"o}rster
  Schreiber}, {Shapiro}, {Genel}, {Lilly}, and {et~al.}]{Genzel2011}
{Genzel}, R.; {Newman}, S.; {Jones}, T.; {F{\"o}rster Schreiber}, N.M.;
  {Shapiro}, K.; {Genel}, S.; {Lilly}, S.J.; {et~al.}.
\newblock {The Sins Survey of z \~{} 2 Galaxy Kinematics: Properties of the
  Giant Star-forming Clumps}.
\newblock {\em \apj} {\bf 2011}, {\em 733},~101,
  \href{http://xxx.lanl.gov/abs/1011.5360}{{\normalfont
  [arXiv:astro-ph.CO/1011.5360]}}.
\newblock
  doi:{\changeurlcolor{black}\href{https://doi.org/10.1088/0004-637X/733/2/101}{\detokenize{10.1088/0004-637X/733/2/101}}}.

\bibitem[{Guo} \em{et~al.}(2012){Guo}, {Giavalisco}, {Ferguson}, {Cassata}, and
  {Koekemoer}]{Guo2012}
{Guo}, Y.; {Giavalisco}, M.; {Ferguson}, H.C.; {Cassata}, P.; {Koekemoer}, A.M.
\newblock {Multi-wavelength View of Kiloparsec-scale Clumps in Star-forming
  Galaxies at z \~{} 2}.
\newblock {\em \apj} {\bf 2012}, {\em 757},~120,
  \href{http://xxx.lanl.gov/abs/1110.3800}{{\normalfont [1110.3800]}}.
\newblock
  doi:{\changeurlcolor{black}\href{https://doi.org/10.1088/0004-637X/757/2/120}{\detokenize{10.1088/0004-637X/757/2/120}}}.

\bibitem[{Wuyts} \em{et~al.}(2013){Wuyts}, {F{\"o}rster Schreiber}, {Nelson},
  {van Dokkum}, {Brammer}, {Chang}, {Faber}, {Ferguson}, {Franx}, {Fumagalli},
  {Genzel}, {Grogin}, {Kocevski}, {Koekemoer}, {Lundgren}, {Lutz}, {McGrath},
  {Momcheva}, {Rosario}, {Skelton}, {Tacconi}, {van der Wel}, and
  {Whitaker}]{Wuyts2013}
{Wuyts}, S.; {F{\"o}rster Schreiber}, N.M.; {Nelson}, E.J.; {van Dokkum}, P.G.;
  {Brammer}, G.; {Chang}, Y.Y.; {Faber}, S.M.; {Ferguson}, H.C.; {Franx}, M.;
  {Fumagalli}, M.;  et~al.
\newblock {A CANDELS-3D-HST synergy: Resolved Star Formation Patterns at 0.7 <
  z < 1.5}.
\newblock {\em \apj} {\bf 2013}, {\em 779},~135,
  \href{http://xxx.lanl.gov/abs/1310.5702}{{\normalfont [1310.5702]}}.
\newblock
  doi:{\changeurlcolor{black}\href{https://doi.org/10.1088/0004-637X/779/2/135}{\detokenize{10.1088/0004-637X/779/2/135}}}.

\bibitem[{Guo} \em{et~al.}(2015){Guo}, {Ferguson}, {Bell}, {Koo}, {Conselice},
  {Giavalisco}, {Kassin}, {Lu}, {Lucas}, {Mandelker}, {McIntosh}, {Primack},
  {Ravindranath}, {Barro}, {Ceverino}, {Dekel}, {Faber}, {Fang}, {Koekemoer},
  {Noeske}, {Rafelski}, and {Straughn}]{Guo2015}
{Guo}, Y.; {Ferguson}, H.C.; {Bell}, E.F.; {Koo}, D.C.; {Conselice}, C.J.;
  {Giavalisco}, M.; {Kassin}, S.; {Lu}, Y.; {Lucas}, R.; {Mandelker}, N.;
  et~al.
\newblock {Clumpy Galaxies in CANDELS. I. The Definition of UV Clumps and the
  Fraction of Clumpy Galaxies at 0.5 < z < 3}.
\newblock {\em \apj} {\bf 2015}, {\em 800},~39,
  \href{http://xxx.lanl.gov/abs/1410.7398}{{\normalfont [1410.7398]}}.
\newblock
  doi:{\changeurlcolor{black}\href{https://doi.org/10.1088/0004-637X/800/1/39}{\detokenize{10.1088/0004-637X/800/1/39}}}.

\bibitem[{Elmegreen} \em{et~al.}(2007){Elmegreen}, {Elmegreen}, {Ravindranath},
  and {Coe}]{Elmegreen2007}
{Elmegreen}, D.M.; {Elmegreen}, B.G.; {Ravindranath}, S.; {Coe}, D.A.
\newblock {Resolved Galaxies in the Hubble Ultra Deep Field: Star Formation in
  Disks at High Redshift}.
\newblock {\em \apj} {\bf 2007}, {\em 658},~763--777,
  \href{http://xxx.lanl.gov/abs/astro-ph/0701121}{{\normalfont
  [astro-ph/0701121]}}.
\newblock
  doi:{\changeurlcolor{black}\href{https://doi.org/10.1086/511667}{\detokenize{10.1086/511667}}}.

\bibitem[{Noguchi}(1998)]{Noguchi1998}
{Noguchi}, M.
\newblock {Clumpy star-forming regions as the origin of the peculiar morphology
  of high-redshift galaxies}.
\newblock {\em \nat} {\bf 1998}, {\em 392},~253.
\newblock
  doi:{\changeurlcolor{black}\href{https://doi.org/10.1038/32596}{\detokenize{10.1038/32596}}}.

\bibitem[{Noguchi}(1999)]{Noguchi1999}
{Noguchi}, M.
\newblock {Early Evolution of Disk Galaxies: Formation of Bulges in Clumpy
  Young Galactic Disks}.
\newblock {\em \apj} {\bf 1999}, {\em 514},~77--95,
  \href{http://xxx.lanl.gov/abs/astro-ph/9806355}{{\normalfont
  [astro-ph/9806355]}}.
\newblock
  doi:{\changeurlcolor{black}\href{https://doi.org/10.1086/306932}{\detokenize{10.1086/306932}}}.

\bibitem[{Aumer} \em{et~al.}(2010){Aumer}, {Burkert}, {Johansson}, and
  {Genzel}]{Aumer2010}
{Aumer}, M.; {Burkert}, A.; {Johansson}, P.H.; {Genzel}, R.
\newblock {The Structure of Gravitationally Unstable Gas-rich Disk Galaxies}.
\newblock {\em \apj} {\bf 2010}, {\em 719},~1230--1243,
  \href{http://xxx.lanl.gov/abs/1007.0169}{{\normalfont [1007.0169]}}.
\newblock
  doi:{\changeurlcolor{black}\href{https://doi.org/10.1088/0004-637X/719/2/1230}{\detokenize{10.1088/0004-637X/719/2/1230}}}.

\bibitem[{Baumgardt} and {Kroupa}(2007)]{Baumgardt2007}
{Baumgardt}, H.; {Kroupa}, P.
\newblock {A comprehensive set of simulations studying the influence of gas
  expulsion on star cluster evolution}.
\newblock {\em \mnras} {\bf 2007}, {\em 380},~1589--1598,
  \href{http://xxx.lanl.gov/abs/0707.1944}{{\normalfont [0707.1944]}}.
\newblock
  doi:{\changeurlcolor{black}\href{https://doi.org/10.1111/j.1365-2966.2007.12209.x}{\detokenize{10.1111/j.1365-2966.2007.12209.x}}}.

\bibitem[{Krumholz} and {Tan}(2007)]{Krumholz2007}
{Krumholz}, M.R.; {Tan}, J.C.
\newblock {Slow Star Formation in Dense Gas: Evidence and Implications}.
\newblock {\em \apj} {\bf 2007}, {\em 654},~304--315.
\newblock
  doi:{\changeurlcolor{black}\href{https://doi.org/10.1086/509101}{\detokenize{10.1086/509101}}}.

\bibitem[{Elmegreen} \em{et~al.}(2008){Elmegreen}, {Bournaud}, and
  {Elmegreen}]{Elmegreen2008}
{Elmegreen}, B.G.; {Bournaud}, F.; {Elmegreen}, D.M.
\newblock {Bulge Formation by the Coalescence of Giant Clumps in Primordial
  Disk Galaxies}.
\newblock {\em \apj} {\bf 2008}, {\em 688},~67--77,
  \href{http://xxx.lanl.gov/abs/0808.0716}{{\normalfont [0808.0716]}}.
\newblock
  doi:{\changeurlcolor{black}\href{https://doi.org/10.1086/592190}{\detokenize{10.1086/592190}}}.

\bibitem[{Elmegreen} \em{et~al.}(2013){Elmegreen}, {Elmegreen}, {S{\'a}nchez
  Almeida}, {Mu{\~n}oz-Tu{\~n}{\'o}n}, {Dewberry}, {Putko}, {Teich}, and
  {Popinchalk}]{Elmegreen2013}
{Elmegreen}, B.G.; {Elmegreen}, D.M.; {S{\'a}nchez Almeida}, J.;
  {Mu{\~n}oz-Tu{\~n}{\'o}n}, C.; {Dewberry}, J.; {Putko}, J.; {Teich}, Y.;
  {Popinchalk}, M.
\newblock {Massive Clumps in Local Galaxies: Comparisons with High-redshift
  Clumps}.
\newblock {\em \apj} {\bf 2013}, {\em 774},~86,
  \href{http://xxx.lanl.gov/abs/1308.0306}{{\normalfont [1308.0306]}}.
\newblock
  doi:{\changeurlcolor{black}\href{https://doi.org/10.1088/0004-637X/774/1/86}{\detokenize{10.1088/0004-637X/774/1/86}}}.

\bibitem[{Garland} \em{et~al.}(2015){Garland}, {Pisano}, {Mac Low}, {Kreckel},
  {Rabidoux}, and {Guzm{\'a}n}]{Garland2015}
{Garland}, C.A.; {Pisano}, D.J.; {Mac Low}, M.M.; {Kreckel}, K.; {Rabidoux},
  K.; {Guzm{\'a}n}, R.
\newblock {Nearby Clumpy, Gas Rich, Star-forming Galaxies: Local Analogs of
  High-redshift Clumpy Galaxies}.
\newblock {\em \apj} {\bf 2015}, {\em 807},~134,
  \href{http://xxx.lanl.gov/abs/1506.04649}{{\normalfont [1506.04649]}}.
\newblock
  doi:{\changeurlcolor{black}\href{https://doi.org/10.1088/0004-637X/807/2/134}{\detokenize{10.1088/0004-637X/807/2/134}}}.

\bibitem[Mandelker \em{et~al.}(2017)Mandelker, Dekel, Ceverino, DeGraf, Guo,
  and Primack]{Mandelker2015}
Mandelker, N.; Dekel, A.; Ceverino, D.; DeGraf, C.; Guo, Y.; Primack, J.
\newblock {Giant Clumps in Simulated High-z Galaxies: Properties, Evolution and
  Dependence on Feedback}.
\newblock {\em Mon. Not. Roy. Astron. Soc.} {\bf 2017}, {\em 464},~635--665,
  \href{http://xxx.lanl.gov/abs/1512.08791}{{\normalfont
  [arXiv:astro-ph.GA/1512.08791]}}.
\newblock
  doi:{\changeurlcolor{black}\href{https://doi.org/10.1093/mnras/stw2358}{\detokenize{10.1093/mnras/stw2358}}}.

\bibitem[{White} and {Frenk}(1991)]{White1991}
{White}, S.D.M.; {Frenk}, C.S.
\newblock {Galaxy formation through hierarchical clustering}.
\newblock {\em \apj} {\bf 1991}, {\em 379},~52--79.
\newblock
  doi:{\changeurlcolor{black}\href{https://doi.org/10.1086/170483}{\detokenize{10.1086/170483}}}.

\bibitem[{Kravtsov} \em{et~al.}(2004){Kravtsov}, {Gnedin}, and
  {Klypin}]{Kravtsov2004}
{Kravtsov}, A.V.; {Gnedin}, O.Y.; {Klypin}, A.A.
\newblock {The Tumultuous Lives of Galactic Dwarfs and the Missing Satellites
  Problem}.
\newblock {\em \apj} {\bf 2004}, {\em 609},~482--497,
  \href{http://xxx.lanl.gov/abs/astro-ph/0401088}{{\normalfont
  [astro-ph/0401088]}}.
\newblock
  doi:{\changeurlcolor{black}\href{https://doi.org/10.1086/421322}{\detokenize{10.1086/421322}}}.

\bibitem[{Croton} \em{et~al.}(2006){Croton}, {Springel}, {White}, {De Lucia},
  {Frenk}, {Gao}, {Jenkins}, {Kauffmann}, {Navarro}, and {Yoshida}]{Croton2006}
{Croton}, D.J.; {Springel}, V.; {White}, S.D.M.; {De Lucia}, G.; {Frenk}, C.S.;
  {Gao}, L.; {Jenkins}, A.; {Kauffmann}, G.; {Navarro}, J.F.; {Yoshida}, N.
\newblock {The many lives of active galactic nuclei: cooling flows, black holes
  and the luminosities and colours of galaxies}.
\newblock {\em \mnras} {\bf 2006}, {\em 365},~11--28,
  \href{http://xxx.lanl.gov/abs/astro-ph/0508046}{{\normalfont
  [astro-ph/0508046]}}.
\newblock
  doi:{\changeurlcolor{black}\href{https://doi.org/10.1111/j.1365-2966.2005.09675.x}{\detokenize{10.1111/j.1365-2966.2005.09675.x}}}.

\bibitem[{Chabrier}(2003)]{Chabrier2003}
{Chabrier}, G.
\newblock {Galactic Stellar and Substellar Initial Mass Function}.
\newblock {\em \pasp} {\bf 2003}, {\em 115},~763--795,
  \href{http://xxx.lanl.gov/abs/astro-ph/0304382}{{\normalfont
  [astro-ph/0304382]}}.
\newblock
  doi:{\changeurlcolor{black}\href{https://doi.org/10.1086/376392}{\detokenize{10.1086/376392}}}.

\bibitem[Booth and Schaye(2009)]{Booth2009}
Booth, C.M.; Schaye, J.
\newblock Cosmological simulations of the growth of supermassive black holes
  and feedback from active galactic nuclei: method and tests.
\newblock {\em Monthly Notices of the Royal Astronomical Society} {\bf 2009},
  {\em 398},~53--74.
\newblock
  doi:{\changeurlcolor{black}\href{https://doi.org/10.1111/j.1365-2966.2009.15043.x}{\detokenize{10.1111/j.1365-2966.2009.15043.x}}}.

\bibitem[{Martizzi} \em{et~al.}(2012){Martizzi}, {Teyssier}, and
  {Moore}]{Martizzi2012a}
{Martizzi}, D.; {Teyssier}, R.; {Moore}, B.
\newblock {The formation of the brightest cluster galaxies in cosmological
  simulations: the case for active galactic nucleus feedback}.
\newblock {\em \mnras} {\bf 2012}, {\em 420},~2859--2873,
  \href{http://xxx.lanl.gov/abs/1106.5371}{{\normalfont
  [arXiv:astro-ph.CO/1106.5371]}}.
\newblock
  doi:{\changeurlcolor{black}\href{https://doi.org/10.1111/j.1365-2966.2011.19950.x}{\detokenize{10.1111/j.1365-2966.2011.19950.x}}}.

\bibitem[{Cattaneo} \em{et~al.}(2006){Cattaneo}, {Dekel}, {Devriendt},
  {Guiderdoni}, and {Blaizot}]{Cattaneo2006}
{Cattaneo}, A.; {Dekel}, A.; {Devriendt}, J.; {Guiderdoni}, B.; {Blaizot}, J.
\newblock {Modelling the galaxy bimodality: shutdown above a critical halo
  mass}.
\newblock {\em \mnras} {\bf 2006}, {\em 370},~1651--1665,
  \href{http://xxx.lanl.gov/abs/astro-ph/0601295}{{\normalfont
  [astro-ph/0601295]}}.
\newblock
  doi:{\changeurlcolor{black}\href{https://doi.org/10.1111/j.1365-2966.2006.10608.x}{\detokenize{10.1111/j.1365-2966.2006.10608.x}}}.

\bibitem[{Klypin} \em{et~al.}(2011){Klypin}, {Trujillo-Gomez}, and
  {Primack}]{Klypin2011}
{Klypin}, A.A.; {Trujillo-Gomez}, S.; {Primack}, J.
\newblock {Dark Matter Halos in the Standard Cosmological Model: Results from
  the Bolshoi Simulation}.
\newblock {\em \apj} {\bf 2011}, {\em 740},~102,
  \href{http://xxx.lanl.gov/abs/1002.3660}{{\normalfont [1002.3660]}}.
\newblock
  doi:{\changeurlcolor{black}\href{https://doi.org/10.1088/0004-637X/740/2/102}{\detokenize{10.1088/0004-637X/740/2/102}}}.

\bibitem[{Benson}(2012)]{Benson2012}
{Benson}, A.J.
\newblock {G ALACTICUS: A semi-analytic model of galaxy formation}.
\newblock {\em \na} {\bf 2012}, {\em 17},~175--197,
  \href{http://xxx.lanl.gov/abs/1008.1786}{{\normalfont [1008.1786]}}.
\newblock
  doi:{\changeurlcolor{black}\href{https://doi.org/10.1016/j.newast.2011.07.004}{\detokenize{10.1016/j.newast.2011.07.004}}}.

\bibitem[{Wang} \em{et~al.}(2015){Wang}, {Dutton}, {Stinson}, {Macci{\`o}},
  {Penzo}, {Kang}, {Keller}, and {Wadsley}]{Wang2015}
{Wang}, L.; {Dutton}, A.A.; {Stinson}, G.S.; {Macci{\`o}}, A.V.; {Penzo}, C.;
  {Kang}, X.; {Keller}, B.W.; {Wadsley}, J.
\newblock {NIHAO project - I. Reproducing the inefficiency of galaxy formation
  across cosmic time with a large sample of cosmological hydrodynamical
  simulations}.
\newblock {\em \mnras} {\bf 2015}, {\em 454},~83--94,
  \href{http://xxx.lanl.gov/abs/1503.04818}{{\normalfont [1503.04818]}}.
\newblock
  doi:{\changeurlcolor{black}\href{https://doi.org/10.1093/mnras/stv1937}{\detokenize{10.1093/mnras/stv1937}}}.

\bibitem[{Moster} \em{et~al.}(2013){Moster}, {Naab}, and {White}]{Moster2013}
{Moster}, B.P.; {Naab}, T.; {White}, S.D.M.
\newblock {Galactic star formation and accretion histories from matching
  galaxies to dark matter haloes}.
\newblock {\em \mnras} {\bf 2013}, {\em 428},~3121--3138,
  \href{http://xxx.lanl.gov/abs/1205.5807}{{\normalfont
  [arXiv:astro-ph.CO/1205.5807]}}.
\newblock
  doi:{\changeurlcolor{black}\href{https://doi.org/10.1093/mnras/sts261}{\detokenize{10.1093/mnras/sts261}}}.

\bibitem[{Behroozi} \em{et~al.}(2013){Behroozi}, {Wechsler}, and
  {Conroy}]{Behroozi2013}
{Behroozi}, P.S.; {Wechsler}, R.H.; {Conroy}, C.
\newblock {The Average Star Formation Histories of Galaxies in Dark Matter
  Halos from z = 0-8}.
\newblock {\em \apj} {\bf 2013}, {\em 770},~57,
  \href{http://xxx.lanl.gov/abs/1207.6105}{{\normalfont
  [arXiv:astro-ph.CO/1207.6105]}}.
\newblock
  doi:{\changeurlcolor{black}\href{https://doi.org/10.1088/0004-637X/770/1/57}{\detokenize{10.1088/0004-637X/770/1/57}}}.

\bibitem[{Behroozi} \em{et~al.}(2019){Behroozi}, {Wechsler}, {Hearin}, and
  {Conroy}]{Behroozi2019}
{Behroozi}, P.; {Wechsler}, R.H.; {Hearin}, A.P.; {Conroy}, C.
\newblock {UNIVERSEMACHINE: The correlation between galaxy growth and dark
  matter halo assembly from z = 0-10}.
\newblock {\em \mnras} {\bf 2019}, {\em 488},~3143--3194,
  \href{http://xxx.lanl.gov/abs/1806.07893}{{\normalfont
  [arXiv:astro-ph.GA/1806.07893]}}.
\newblock
  doi:{\changeurlcolor{black}\href{https://doi.org/10.1093/mnras/stz1182}{\detokenize{10.1093/mnras/stz1182}}}.

\bibitem[{Planck Collaboration} \em{et~al.}(2014){Planck Collaboration}, {Ade},
  {Aghanim}, {Armitage-Caplan}, {Arnaud}, {Ashdown}, {Atrio-Barandela},
  {Aumont}, {Baccigalupi}, {Banday}, and et~al.]{Planck2014}
{Planck Collaboration}.; {Ade}, P.A.R.; {Aghanim}, N.; {Armitage-Caplan}, C.;
  {Arnaud}, M.; {Ashdown}, M.; {Atrio-Barandela}, F.; {Aumont}, J.;
  {Baccigalupi}, C.; {Banday}, A.J.;  et~al.
\newblock {Planck 2013 results. XVI. Cosmological parameters}.
\newblock {\em \aap} {\bf 2014}, {\em 571},~A16,
  \href{http://xxx.lanl.gov/abs/1303.5076}{{\normalfont [1303.5076]}}.
\newblock
  doi:{\changeurlcolor{black}\href{https://doi.org/10.1051/0004-6361/201321591}{\detokenize{10.1051/0004-6361/201321591}}}.

\end{thebibliography}
\end{adjustwidth}
\end{document}